%% file: main.tex
\documentclass[twocolumn]{article}

\usepackage{graphicx}
\usepackage{tikz}
\usetikzlibrary{arrows, shapes.geometric, positioning, shadows, calc, fit, arrows.meta, backgrounds}
\usepackage{booktabs}
\usepackage{amssymb}
\usepackage{amsmath}
\usepackage{algorithm}
\usepackage[breaklinks=true]{hyperref}
\usepackage{xurl}
\usepackage[numbers]{natbib}
\usepackage{microtype}
\usepackage{geometry}
\usepackage{xcolor}
\usepackage{xspace}
\usepackage{stfloats}
\usepackage{float}
\usepackage{listings}
\usepackage{caption}
\usepackage{pgfplots}
\pgfplotsset{compat=1.18}

\newcommand{\paperOneCorpus}{the locked LLM corpus\xspace}
\newcommand{\webCorpus}{the locked Web corpus\xspace}

\begin{document}

\title{From Attack Simulation to SIEM Rule:\\
Deterministic Detection-as-Code Synthesis\\
with Probe-Level Traceability}

\author{Alexandre Cristov\~ao Maiorano\\
\texttt{alexandre@lumytics.com}}
\date{}

\twocolumn[
  \begin{@twocolumnfalse}
    \maketitle
    \input{sections/abstract}
    \vspace{2mm}
  \end{@twocolumnfalse}
]

\input{sections/intro}
\input{sections/related}

\input{sections/methodology}
\input{sections/results}

\input{sections/discussion}
\input{sections/conclusion}

\section*{AI Tools Disclosure}

This research leveraged AI-assisted development tools to support
manuscript preparation and code development, while maintaining full
human oversight and accountability. The following tools were used:
\begin{itemize}
    \item \textbf{Language models:} GPT-5 family (OpenAI via Codex)
    and Claude Opus 4.7 (Anthropic Claude Code) were used to generate
    and review code implementations, and to refine manuscript text.
    Google Gemini 2.5 Flash is also a \emph{measurement subject} in
    this paper: it is the LLM backend of the target application that
    the probes are fired against (Section~\ref{sec:methodology}); that
    usage is part of the system being measured and is not an authoring
    use. The rule synthesis itself is deterministic and uses no model.

    \item \textbf{Web search:} MCP Tavily integration was used to
    support literature review and fact-checking during manuscript
    preparation.
\end{itemize}
All scientific arguments, empirical methodology, statistical
analysis, research questions, and conclusions were independently
conceived, developed, and validated by the author.

\section*{Statements and Declarations}
\paragraph{Funding}
This study did not receive a direct research grant. Experimental
operation used Google Cloud resources and Gemini 2.5 Flash as the
LLM target backend.

\paragraph{Competing Interests}
The author declares no competing interests.

\paragraph{Data Availability}
All aggregate results and statistical tables are reported in full
in the paper (Section~\ref{sec:results}); every numeric value in
those tables traces back to a JSON artifact in the public
replication package
(\url{https://github.com/alemaiorano/detection-as-code-synthesis}).
The released artifacts are: the two locked probe corpora
(\texttt{owasp-llm-probe-corpus.json}, \texttt{juiceshop-probes.json}),
the synthesized rule sets and per-run records, the OpenSearch replay
queries, the sha256-pinned held-out AdvBench and HarmBench subsets
(with the build scripts that regenerate them from upstream), and the
multi-seed calibration outputs. A \texttt{CLAIMS\_TO\_ARTIFACTS.csv}
maps every reported number to its artifact field and the script that
regenerates it.

\emph{Withheld for intellectual-property reasons:} the production
source of the BAS engine that fires the probes and emits the
\texttt{Finding} records is proprietary, as is the production source
of the synthesis function and template internals. \textbf{However,
neither is required to reproduce any number in this paper.} The
synthesis function --- the contribution --- is published as
language-agnostic pseudocode (the full
\texttt{pickTemplate}/\texttt{findingToSigma} contract) sufficient to
re-implement it; it is pure on \texttt{(finding, template)}, so any
verdict-producing engine that emits the finding fields in
Section~\ref{sec:methodology:findings} can drive it, and the held-out
and real-SIEM measurements consume the \emph{emitted rules} (re-derivable
from the corpora and the specified synthesis) and the public
AdvBench/HarmBench prompt sets --- never the engine. The LLM-target
backend is Gemini 2.5 Flash (commercial API); independent replication
requires a Google Cloud account but no proprietary tooling.

\paragraph{Code Availability}
The deterministic synthesis function is specified as language-agnostic
pseudocode (\texttt{synthesis/SYNTHESIS\_SPEC.md}), alongside the
template-library metadata, the locked corpora, and the standalone
exporters that regenerate every table and figure from the artifacts,
in the public replication package
(\url{https://github.com/alemaiorano/detection-as-code-synthesis}),
under CC-BY 4.0. The production engine source remains proprietary but
is not required to reproduce the reported numbers
(see the package's \texttt{SCOPE.md}).

\paragraph{Responsible Research and Dual-Use}
This work is defensive: its output is detection content (Sigma rules)
that helps SOC teams catch attacks they would otherwise miss. The
released artifacts introduce no new offensive capability. The attack
probes target only intentionally-vulnerable applications we ship for
the purpose (OWASP Juice Shop and a deliberately-insecure VulnBot);
any secrets in the corpora are clearly-labelled non-functional
canaries, not real credentials. The held-out evaluation reuses
AdvBench~\cite{zouAdvBench} and HarmBench~\cite{harmbench}, established
public benchmarks redistributed only as sha256-pinned subsets under
their upstream licenses; we add no novel harmful prompts. Simulation
is to be run only against systems the operator owns or is authorised
to test.

\paragraph{Ethics Approval and Consent to Participate}
Not applicable: the study involves no human subjects, personal data,
or live third-party systems --- only the authors' own sandbox targets.

\paragraph{Consent for Publication}
Not applicable.

\bibliographystyle{plainnat}
\bibliography{references}

\input{sections/appendix}

\end{document}

%% file: sections/abstract.tex
\begin{abstract}
Security teams routinely simulate attacks against their own systems
to check whether their monitoring would catch a real intruder. These
Breach-and-Attack-Simulation (BAS) tools surface findings, but the
security information and event management (SIEM) systems that watch
production need detection rules --- and today a human bridges that gap
by hand, reading each finding and writing the corresponding Sigma rule
(a vendor-neutral detection format). We show this translation can be
partially automated when probes are drawn from a \emph{locked corpus},
so each finding carries a stable identifier back to the originating
probe.
We describe a deterministic synthesis function that maps each
finding to a starter Sigma rule through a small template library
($N{=}23$, indexed by categories from the OWASP LLM and Web Top 10),
with a back-reference to the originating finding and its MITRE
ATT\&CK technique. On two locked corpora (17-probe LLM, 23-probe
Web), every bypassed-probe finding yields a starter rule, and all
$17/17$ emitted rules parse and convert to Splunk and Elasticsearch
backends. Replayed through a live OpenSearch SIEM, the LLM rules
fire on $30\%$ of a held-out AdvBench subset and $14\%$ of HarmBench
at $7.7\%$ false positives on a benign baseline; the Web side is
validated structurally, not against a held-out attack set.
The contribution is a verifiable, byte-stable path from BAS finding
to operator-deployable starter rule, re-derivable from the published
corpus and template library alone --- trading the breadth of
LLM-generative methods for exact reproducibility and a typed traceback
from any fired alert to the originating probe.
\end{abstract}

%% file: sections/intro.tex
\section{Introduction}

A typical BAS run produces dozens to hundreds of findings. Each finding
describes \emph{what} the simulator was able to do against the target.
The next step in a security operations center (SOC) workflow is to
ask: \emph{would we have detected this in production?} — and if the
answer is no, write a detection rule, typically in the Sigma
format~\cite{sigmaHQ} — a vendor-neutral YAML schema for detection
rules — so it can be ported across security information and event
management (SIEM) backends.

Today that translation is manual. The finding's textual description
(\texttt{detection\_gap}, \texttt{defense\_recommendation}) gives the
analyst enough context to author a rule, but they still have to choose
the data source, write the regex/keyword logic, attach the right MITRE
ATT\&CK tag, and decide on a sensitivity threshold. The result is
craftwork: high quality when the analyst is good, but slow, expensive,
and dependent on which analyst sat down with the finding.

Two recent developments in the BAS-research literature change the
shape of the problem. First, the move toward \emph{locked probe
corpora} — published JSON files whose every entry has a stable
\texttt{probe\_id} that does not change between runs~\cite{paperOne}.
Second, the move toward OWASP-aligned
categorization~\cite{owaspLlmTop10} --- OWASP being the community body
whose Top 10 lists enumerate the most critical security risks for web
and LLM applications --- where each probe declares its OWASP LLM Top
10 (or Web Top 10) category up front and tags itself with the
corresponding technique from MITRE ATT\&CK~\cite{mitreAttack}, the
standard public knowledge base of adversary tactics and techniques.
These
two together mean a finding now carries enough structure to look up a
template-shaped rule deterministically:
\texttt{finding.metadata.probe\_id} $\to$ corpus entry $\to$
OWASP category $\to$ Sigma template.

This paper contributes the following:

\begin{enumerate}
\item A \textbf{template library} of 23 Sigma rule skeletons (8 MITRE
T-code legacy templates + 9 OWASP LLM Top 10 + 6 OWASP Web Top 10;
19 exercised by the two released corpora, 4 reserved for future
corpora covering A04, LLM03, LLM04, LLM05) that resolve findings
emitted by the locked corpora to starter detection rules. The library
metadata and the synthesis contract that consumes it are published as
pseudocode in the replication package, under the same license as the
corpora.

\item A \textbf{traceability contract}: every emitted rule contains
two reference URIs — one to the originating finding and one to the
MITRE technique. Combined with the finding's stable
\texttt{probe\_id} and a content-hashed record of the run that
produced it, this gives a reviewer a deterministic path from rule
back to the corpus entry that motivated it.

\item A \textbf{reproducibility envelope}: an external reviewer can
re-derive every emitted rule from the published locked corpus, the
published template library, and the deterministic synthesis
function. The protocol is detailed in
Section~\ref{sec:methodology:reproducibility}; we do not require the
reader to run our specific software stack to verify any claim.

\item An \textbf{operational measurement} of coverage on the two
released corpora plus a real-SIEM replay: across 17 LLM probes and
23 Web probes, every bypassed-probe finding produced a starter rule
(skipped=0). The emitted rules ingest into a live
OpenSearch~\cite{opensearch} index via the Lucene backend and fire
on $15/50$ of a held-out AdvBench~\cite{zouAdvBench} subset and
$7/50$ of a second held-out HarmBench~\cite{harmbench} subset, with
$7.7\%$ FP on a benign-LLM baseline. The measurement protocol is
detailed in Section~\ref{sec:methodology} and the results are
reported in Sections~\ref{sec:results:heldout}
and~\ref{sec:results:opensearch}.
\end{enumerate}

\paragraph{From v1 to v2: rubric evolution.} The version labels v1 and
v2 in this paper refer only to the template-authoring rubric --- how a
template's detection logic is written --- and not to the engine's
feature scope, which is held fixed throughout. The generalization
results in contribution~(4) reflect an iterative improvement of that
rubric. Our first template rubric (v1) used keyword-only matching and
achieved a
$0/50$ fire rate on the AdvBench held-out set — a baseline failure
that motivated a v2 authoring rubric adding regex selection on
semantic markers. A Python prototype of the v2 rubric achieved
$31/50$ ($62\%$); after integration into the engine, the v2 rubric
achieved $30/50$ ($60\%$) in live synthesis, and $15/50$ ($30\%$)
when replayed through a real OpenSearch SIEM (the drop attributable to
a backend compatibility shim, as discussed in
Section~\ref{sec:discussion:threats}). The numbers reported in this
paper therefore represent a progression: v1 (keyword-only, $0/50$),
v2-prototype (Python, $31/50$), v2-integrated (engine, $30/50$), and
v2-integrated (SIEM, $15/50$). Section~\ref{sec:results} provides a
reading guide that disambiguates each measurement context.

Why this matters beyond the immediate time saving: determinism and
probe-level traceability are what let auto-generated detection content
be \emph{governed} like code --- version-controlled, diffed across
corpus revisions, reviewed, and signed off with an unambiguous owner ---
rather than treated as opaque output a SOC must re-verify on every run.
That is the discipline the ``detection-as-code'' movement asks for, and
it is precisely what generative approaches, whose output changes with
the model release or prompt, cannot guarantee. A reproducible
finding-to-rule path is therefore a prerequisite for trusting
automation in detection engineering, not merely a convenience.

This paper is positioned in the gap between \emph{BAS surface
results} and \emph{ready-to-deploy SIEM content}. We do not claim
that auto-generated rules are production-ready. The system is
\emph{designed} to shorten the manual-authoring loop by giving
the analyst a typed, referenceable starting point with the corpus
probe baked in; we do not run a user study or a time-on-task
trial in this paper, so the workflow speed-up claim is supported
structurally (per-rule synth time of milliseconds plus a few
minutes of analyst review, against a 30--60-minute manual
baseline; break-even at $\sim$4--8 findings per category, see
Section~\ref{sec:discussion:authoring}) rather than empirically.
A controlled user study is the natural follow-up.

% Bib references resolved in references.bib — paperOne points to the
% companion LLM-BAS-Coverage paper.

%% file: sections/related.tex
\section{Related Work and Positioning}
\label{sec:related}

Four bodies of work are relevant. We summarize positioning in
Table~\ref{tab:sota} and elaborate below.

\paragraph{BAS calibration with locked corpora.} Our companion
work~\cite{paperOne} establishes the per-defense attribution
methodology that produces the findings consumed here. We extend
the corpus-locked discipline from the engine side
(probe~$\to$~finding) to the defense side (finding~$\to$~rule):
the same sha256-pinned probe corpus that anchors that paper's
attribution measurements anchors our rule synthesis, and the
synthesis function is a pure mapping from (finding, template) to
Sigma YAML.

\paragraph{LLM red-team frameworks.} garak~\cite{garak} is the
closest peer for the LLM-side calibration scenario: it ships
locked probes keyed to OWASP LLM Top 10 + MITRE ATT\&CK and produces
per-probe verdicts. garak does \emph{not} emit detection content;
the operator reads a result table and writes rules manually. Our
synthesis function is the missing back-half — given garak-style
findings, emit Sigma rules deterministically.

\paragraph{SIEM rule libraries (hand-authored).}
SigmaHQ~\cite{sigmaHQrules} maintains $\sim$3{,}000 community-authored
Sigma rules. The library is excellent but the authoring workflow is
exactly the manual step we automate: a human reads a vuln advisory or
red-team report, writes a Sigma YAML, and submits a PR. We do not
compete with SigmaHQ's depth (their rules cover Windows kernel
events, AWS CloudTrail audit, etc.); we offer a starter-rule path
for BAS findings that SigmaHQ does not currently cover (corpus
probe~$\to$~rule with traceback URI).

\paragraph{Adversary emulation platforms.}
CALDERA~\cite{caldera} provides ATT\&CK-mapped adversary abilities
with metadata that detection engineers consume as input to the
manual rule-authoring workflow. Our pipeline differs in two
dimensions: (a) we operate on BAS findings (post-attack signal) not
attack-plan metadata (pre-attack intent); (b) we emit detection
artifacts directly, not just metadata for analysts.

\paragraph{LLM-assisted detection engineering.}
A growing concurrent line uses LLMs as the engine for detection
content. RuleGenie~\cite{ruleGenie} uses LLMs to \emph{optimize}
existing SIEM rules --- tuning thresholds, reducing false
positives. RAM~\cite{ram-arxiv-2025} uses LLMs to map detection
rules to MITRE ATT\&CK techniques. Both produce useful output but
share a property our work deliberately avoids: the mapping from
input to rule (or input to mapping) is not byte-stable across
runs --- a different LLM release, a different prompt, or a
temperature change produces a different artifact for the same
input.

The trade-off is direct. Our deterministic
template-synthesis approach gives up the breadth an LLM gets for
free (an LLM can draft a rule for any attack class given a
prompt; our function returns \texttt{null} on a missing template)
in exchange for two properties LLMs cannot offer: \emph{(i)}
exact reproducibility --- re-running the synthesis on the same
finding emits the same YAML, which a reviewer can audit and a
detection engineer can sign off on without re-verifying each
output; \emph{(ii)} a typed traceability hook ---
every emitted rule carries the originating
\texttt{finding\_id}, so when a rule fires in production the
analyst can pivot back to the BAS probe that produced it, and
when a rule is wrong the owner is unambiguous (the template
author, named in git). Hybrid pipelines that use the
deterministic synthesis as the bootstrapping step and an LLM
optimizer (RuleGenie-style) as the tuning step combine the two
strengths and are the natural composition, not a competing
design.

\input{tables/sota}

%% file: tables/sota.tex
% AUTO-GENERATED comparison table; columns reflect dimensions an
% operator cares about when deciding whether an artifact slots into the
% BAS-finding → SIEM-rule pipeline.
\begin{table*}[t]
  \centering
  \footnotesize
  \caption{Positioning vs.\ closest related artifacts. \emph{Locked corpus} = stable probe IDs survive engine revisions. \emph{Probe traceback} = a rule (or report row) carries an explicit URI back to the originating probe. \emph{Auto-emits Sigma} = a deterministic function emits rule YAML (not a human-authored ruleset). \emph{Multi-backend} = same rules consumable by $\geq 2$ SIEM dialects. \emph{Held-out test} = published, peer-reviewed corpus exercises the artifact outside its training set.}
  \label{tab:sota}
  \resizebox{\textwidth}{!}{%
  \begin{tabular}{lccccc}
    \toprule
    Artifact & Locked corpus & Probe traceback & Auto-emits Sigma & Multi-backend & Held-out test \\
    \midrule
    SigmaHQ rules~\cite{sigmaHQrules}    & N/A & N/A           & no (hand-authored)   & yes (pysigma)  & no \\
    garak~\cite{garak}                   & yes & per-probe id  & no                   & N/A            & partial (corpus is the test) \\
    MITRE CALDERA~\cite{caldera}         & yes (abilities)& per-ability id & no            & N/A            & no \\
    RuleGenie~\cite{ruleGenie}           & no  & N/A           & no (optimizes existing) & N/A         & no \\
    Paper \#1 engine~\cite{paperOne}     & yes & per-probe id  & no (no rule output)   & N/A           & lattice-internal only \\
    \textbf{This work}                   & \textbf{yes (2+2 held-out)} & \textbf{per-rule URI} & \textbf{yes (det.)} & \textbf{yes (Splunk+Lucene)} & \textbf{yes (AdvBench+HarmBench)} \\
    \bottomrule
  \end{tabular}%
  }
\end{table*}

%% file: sections/methodology.tex
\section{Methodology}
\label{sec:methodology}

\input{sections/figure_pipeline}

Figure~\ref{fig:pipeline} shows the round-trip the rest of this
section formalizes: a locked corpus drives the BAS engine --- any
tool that fires each probe against a target application and labels the
outcome as bypassed or blocked --- each bypassed probe emits a
\texttt{Finding} record, the synthesis function converts the finding
to a Sigma rule, and the rule's \texttt{references[0]} URI lets the
SOC analyst click from a fired alert back to the originating probe.
The engine is not the contribution and is deliberately treated as
replaceable: the synthesis function downstream of it is a pure mapping
from (finding, template), so the reproducibility envelope
(Section~\ref{sec:methodology:reproducibility}) holds for any
verdict-producing engine, and the reference implementation we release
is one interchangeable choice.

\subsection{Locked corpora}
\label{sec:methodology:corpora}

\paragraph{Definition.} We use \emph{locked probe corpus} to mean a
JSON file with three properties: (a) every probe carries a stable
\texttt{probe\_id} that survives engine refactors and corpus
revisions; (b) the file content is content-hashed at lock time, and
the hash is part of any claim made against the corpus
(Section~\ref{sec:methodology:reproducibility});
(c) updates to a probe (payload changes, indicator changes) require
a new \texttt{probe\_id} suffix (e.g.\ \texttt{a03-sqli-login-email-001}
$\to$ \texttt{a03-sqli-login-email-002}); the old id is preserved
for backward traceability. The concept is introduced in the
companion BAS-attribution work~\cite{paperOne} and adopted here
without modification. Update policy: a corpus version bump
(\texttt{1.0.0} $\to$ \texttt{1.1.0}) preserves all existing
probe\_ids and adds new ones; a major bump (\texttt{1.x.x} $\to$
\texttt{2.0.0}) is required to retire or repurpose an existing id.

We work with two probe corpora, both released with the artifact pack:

\begin{itemize}
\item \textbf{\paperOneCorpus} — 17 probes covering OWASP LLM01, LLM02,
LLM06, LLM07, LLM10. Locked at \texttt{2026-05-16} for the companion
paper~\cite{paperOne}.
\item \textbf{\webCorpus} — 23 probes covering OWASP A01, A02, A03, A05,
A07, A09 of the OWASP Top 10 (2021)~\cite{owaspTop10}. Locked at
\texttt{2026-05-18}.
\end{itemize}

Each corpus is a JSON file with a schema covering corpus-level
metadata (id, version, lock date) and a list of probes carrying their
\texttt{probe\_id}, OWASP category, target endpoint, payload, expected
indicator, severity, and CWE reference. A content hash of each file
is part of the corpus's lock metadata and is included in the
replication package.

\subsection{Finding emission}
\label{sec:methodology:findings}

A calibration run drives every probe in the corpus against a sandbox
target --- OWASP Juice Shop for the Web corpus, and VulnBot, a
deliberately-vulnerable LLM application we release with this paper
(backed by Gemini 2.5 Flash, model id \texttt{gemini-2.5-flash}, at
sampling temperature $0.7$ --- the model the engine targets in
production, chosen for its cost and latency profile), for the LLM
corpus --- and classifies the response. Probes that bypass
the target's defenses produce a \texttt{Finding} record with the
following fields relevant to detection-rule synthesis:

\begin{itemize}
\item \texttt{attack\_technique} — \texttt{OWASP <CAT> --- <name>}
(e.g.\ \texttt{OWASP LLM01 --- DAN-style act-as hijack}). The OWASP
prefix is the key the template library looks up.
\item \texttt{detection\_gap}, \texttt{defense\_recommendation} —
free-text fields used as Sigma \texttt{description} content.
\item \texttt{metadata.\{corpus\_id, corpus\_version, probe\_id,
owasp\_category\}} — the traceability key.
\item \texttt{evidence\_refs[]} — one entry per run, pointing to a
content-hashed evidence record of the probe-verdict snapshot
(Section~\ref{sec:methodology:reproducibility}).
\end{itemize}

\subsection{Template library and rule synthesis}

The template library contains $N{=}23$ entries indexed by either MITRE
T-code or OWASP category. Lookup order: T-code first (some findings
are labelled with a MITRE ATT\&CK T-code rather than an OWASP
category), OWASP regex second (the calibration scenarios used here). Each template specifies a Sigma
\texttt{logsource} + \texttt{detection} block; it is parameterized
only via the finding's per-rule fields (Listing~\ref{lst:synth-inner}):

\begin{figure*}[tb]
\begin{lstlisting}[basicstyle=\ttfamily\footnotesize,frame=single,framerule=0.4pt,framesep=4pt,linewidth=\textwidth,breaklines=false]
function findingToSigma(finding) {
  template = pickTemplate(finding.attack_technique);
  if (!template) return null;
  return {
    title:       template.title + " --- " + finding.attack_technique,
    id:          finding.finding_id + "-" + template.technique,
    references: [
      "bas://finding/" + finding.finding_id,
      "https://attack.mitre.org/techniques/" + template.technique.replace(".", "/") + "/",
    ],
    tags:       ["attack." + template.technique.toLowerCase(),
                 "bas.severity." + finding.severity],
    logsource:   template.logsource,
    detection:   template.detection,
    falsepositives: template.falsepositives ?? ["Unknown"],
    level:       template.level,
  };
}
\end{lstlisting}
\captionof{lstlisting}{Synthesis is a pure function of (finding, template).}
\label{lst:synth-inner}
\end{figure*}

The synthesis is deterministic: re-running against the same finding
produces the same YAML byte-for-byte (modulo the synthesis date in
\texttt{date:}). This is intentional — the rules can be diffed across
corpus versions to surface template churn.

\paragraph{Template selection logic.} \texttt{pickTemplate(technique)}
applies two passes in order: (1) extract any MITRE ATT\&CK T-code in
the technique string and look it up (e.g.\
\texttt{T1190.001} hits a prompt-injection template); (2) if no T-code
matches, search for an \texttt{OWASP\,<CAT>} prefix in the technique
and look up the OWASP-keyed template (e.g.\ \texttt{LLM01},
\texttt{A03}). When multiple templates would match (rare in practice
— corpus probes carry exactly one OWASP category), the first key tried
wins. This is a first-match heuristic, not a principled selection;
we did not observe ambiguity on the two released corpora but flag two
known weaknesses: (a) a finding whose \texttt{attack\_technique}
string carries both a T-code and an OWASP prefix takes the T-code
path even when the OWASP-keyed template would yield a more specific
rule; (b) cross-cutting attack classes (e.g.\ an SSRF --- server-side request
forgery --- probe that straddles OWASP A01 and A10) cannot map to two
templates in the current pipeline. A principled extension would either treat the
mapping as set-valued (emit multiple rules) or score templates
against the finding's full context (severity, asset, payload shape)
rather than the technique prefix alone. We leave both as future
work.

\paragraph{Finding details that feed into the rule.}
The \texttt{bas://} URI scheme and the \texttt{bas.}-prefixed
tags used throughout this section are the engine's
back-reference convention: \texttt{bas://finding/<id>} resolves
through the engine's evidence-pack API to the originating probe
payload, and \texttt{bas.severity.<lvl>} is a Sigma-compatible
custom tag namespace. Implementations are free to substitute
their own prefix (e.g.\ a tenant-specific URN);
the contract is structural (one URI per finding, one severity
tag per rule), not the literal prefix.
Beyond the template body, four finding fields shape the output: (a)
\texttt{finding\_id} becomes the rule's \texttt{id} (prefixed with
the template's MITRE technique); (b) \texttt{severity} becomes a
\texttt{bas.severity.<lvl>} tag (the Sigma \texttt{level}
field comes from the template, not the finding, since severity in
the corpus is per-probe whereas Sigma \texttt{level} is per-rule);
(c) \texttt{detection\_gap} + \texttt{defense\_recommendation} are
embedded into the rule's \texttt{description} so the analyst sees the
probe evidence inline; (d) the finding's \texttt{metadata.probe\_id}
is encoded in the \texttt{references[0]} URI for traceback.
Crucially, the probe's \emph{payload} is \emph{not} fed in directly
— the template's \texttt{detection} block describes the
\emph{class} of attack (e.g.\ ``\texttt{UNION SELECT} keyword on a
web-server log''), not the specific bytes of the calibration
probe. This decouples the published rule from the calibration probe
distribution and is the design choice that enables future replay-
on-held-out-probes measurement (Section~\ref{sec:methodology:future}).

\paragraph{Traceability URI resolution.} The \texttt{references[0]}
field of each rule encodes a \texttt{finding\_id}. The mapping from
\texttt{finding\_id} to its enclosing run record (which carries
\texttt{probe\_id} and evidence-blob hash) is part of the replication
package. A stand-alone resolver service that lets a SIEM-side rule
click straight through to the originating finding is out of scope
for this paper.

\subsection{Reproducibility envelope}
\label{sec:methodology:reproducibility}

A reviewer should be able to reproduce every emitted rule. The
envelope has four steps, none of which require access to a specific
implementation stack:

\begin{enumerate}
\item Obtain the locked corpus and verify its content hash against
the value committed at lock time.
\item Drive every probe in the corpus through any BAS engine that
produces a per-probe verdict. The method depends on no specific
engine; the production one is withheld, but the synthesis downstream
is engine-agnostic.
\item For each probe whose verdict is ``bypassed,'' construct a
\texttt{Finding} record (fields listed in
Section~\ref{sec:methodology:findings}).
\item Apply the synthesis function --- specified as language-agnostic
pseudocode in the replication package --- to each finding. The
synthesis is deterministic: re-running it on the same finding emits
byte-identical YAML modulo the synthesis-date field. The published
rule set --- each rule's deterministic UUID and its Splunk\,SPL and
Lucene conversions for all $17$ emitted rules --- ships in
\texttt{reports/sigma-conformance*.json}. An independent
re-implementation from the specification must reproduce these:
matching ids (the id is a pure hash of \texttt{(finding\_id,
technique)}) and matching backend conversions, up to YAML
serialization.
\end{enumerate}

The integrity chain is: \emph{locked corpus
$\to$ probe verdict $\to$ \texttt{Finding} record
$\to$ Sigma rule reference}. Each arrow is verifiable from
public artifacts; any break (mismatched corpus hash, missing
\texttt{probe\_id}, dangling finding reference) is detectable
without access to the engine's source code. The corpora, the
synthesis function, every JSON artifact, and a
\texttt{CLAIMS\_TO\_ARTIFACTS.csv} mapping each reported number to its
regenerating script are released as a public replication package
(\url{https://github.com/alemaiorano/detection-as-code-synthesis};
see Data Availability).

\subsection{Measurement protocol}
\label{sec:methodology:future}

Five measurements are relevant: template-hit, FP rate on a benign
baseline, TP rate on a held-out attack corpus, multi-backend Sigma
conformance, and a real-SIEM replay through a live query engine.

\begin{itemize}
\item \textbf{Template-hit rate} (§\ref{sec:results:coverage}) —
percentage of bypassed-probe findings for which the synthesis function
returns a non-null rule. Lower bound on usefulness; upper bound on
what template-only synthesis can achieve.
\item \textbf{Baseline FP rate} (§\ref{sec:results:reproducibility})
— each emitted rule is evaluated against $N{=}100$ benign log lines
(typical product routes for Web, neutral chat prompts for LLM).
Evaluator: keyword/selection substring matching, the v1 template
subset's semantics. This is a synthetic-log harness, not a SIEM.
\item \textbf{Held-out TP rate}
(§\ref{sec:results:heldout}) — each LLM01-family rule is evaluated
against a 50-prompt deterministic sample of
AdvBench~\cite{zouAdvBench}, a public peer-reviewed harmful-prompt
corpus whose contents were never seen at template-authoring time.
A fire is generalization evidence.
\item \textbf{Multi-backend Sigma conformance}
(§\ref{sec:results:coverage}, Table~\ref{tab:v6}) — each emitted
rule is parsed via pysigma and converted to Splunk SPL (Search
Processing Language) and to Elasticsearch's Lucene/KQL (Kibana Query
Language). Reports per-rule parse and conversion success.
\item \textbf{Real-SIEM replay} (§\ref{sec:results:opensearch}) —
each emitted rule is loaded into a standalone OpenSearch index via
its Lucene KQL conversion and queried against the held-out and
benign cohorts. Reports the operator-facing union match rate.
\end{itemize}

%% file: sections/figure_pipeline.tex
% Figure 1 — Linear pipeline diagram for §3 Methodology.
% Six nodes left-to-right with the references[0] back-edge curving
% below the main flow (no overlap with the solid forward arrows).
% Shape discriminates artifact (rounded) vs process (sharp) so the
% figure is legible on B&W print where pastel fills collapse.
\begin{figure*}[t]
  \centering
  \begin{tikzpicture}[
    node distance = 6mm,
    every node/.style = {font=\footnotesize},
    artifact/.style  = {draw=blue!55!black, thick, rounded corners=3pt,
                        fill=blue!12, align=center, inner sep=4pt,
                        minimum height=10mm, minimum width=22mm},
    process/.style   = {draw=orange!70!black, thick,
                        fill=orange!18, align=center, inner sep=4pt,
                        minimum height=10mm, minimum width=22mm},
    flow/.style      = {-{Latex[length=2mm]}, thick},
    trace/.style     = {-{Latex[length=2mm]}, dashed, thick,
                        draw=gray!55!black},
  ]
    \node[artifact]                              (corpus)  {Locked\\corpus};
    \node[process,  right=of corpus]             (engine)  {BAS\\engine};
    \node[artifact, right=of engine]             (finding) {\texttt{Finding}\\\scriptsize\texttt{probe\_id}};
    \node[process,  right=of finding]            (synth)   {\texttt{findingTo}\\\texttt{Sigma()}};
    \node[artifact, right=of synth]              (rule)    {Sigma\\rule};
    \node[process,  right=of rule]               (siem)    {pysigma\\$\to$ SIEM};

    \draw[flow] (corpus)  -- (engine);
    \draw[flow] (engine)  -- (finding);
    \draw[flow] (finding) -- (synth);
    \draw[flow] (synth)   -- (rule);
    \draw[flow] (rule)    -- (siem);

    % Traceability back-edge: rule -> finding via references[0] URI.
    % Curve below the row so it never crosses the solid forward arrows.
    \draw[trace] (rule.south) to[out=-90, in=-90, looseness=1.2]
      node[midway, below, font=\scriptsize, gray!55!black]
        {\texttt{references[0]} URI (\S\ref{sec:methodology:findings})}
      (finding.south);
  \end{tikzpicture}
  \caption{Round-trip from locked corpus to SIEM. Solid arrows: data flow at calibration time (left to right). Dashed arrow: the traceability URI a SOC analyst clicks from a fired rule back to the originating probe. Rounded blue boxes are \emph{artifacts} (the locked corpus, the \texttt{Finding} record, the emitted Sigma rule); sharp orange boxes are \emph{processes} (the BAS engine, the synthesis function, the Sigma backend stack); the right-most box collapses pysigma + Splunk SPL / Lucene KQL conversion + the OpenSearch index that hosts the held-out cohorts (AdvBench, HarmBench, benign-LLM, benign-Web) used in \S\ref{sec:results:opensearch}. The synthesis function is pure on (finding, template) so a reviewer can re-derive the rule set from the leftmost artifacts alone without running the engine source code (\S\ref{sec:methodology:reproducibility}).}
  \label{fig:pipeline}
\end{figure*}
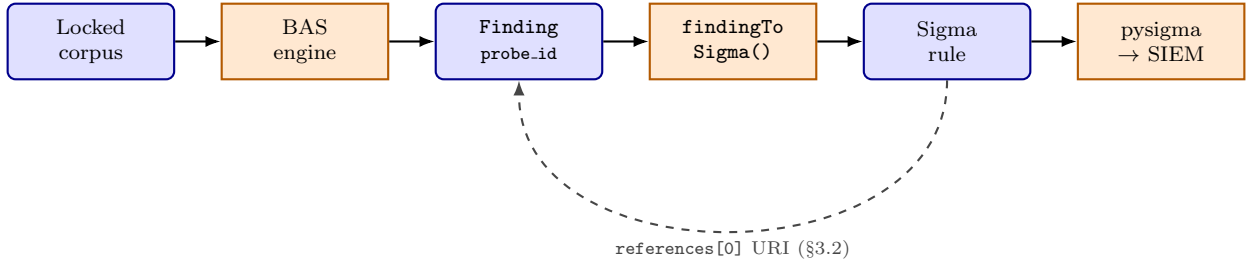

%% file: sections/results.tex
\section{Results}
\label{sec:results}

\paragraph{Reading guide.} The held-out AdvBench corpus appears in
four contexts with four different fire rates ($0/50$, $31/50$,
$30/50$, $15/50$). These are not contradictory: they are one rule set
under progressively stricter evaluation as it evolved from the v1 to
the v2 rubric. \textbf{Table~\ref{tab:v6}(b) collects all four in one
place} and Figure~\ref{fig:firerate} plots them as a single
progression; together they are the canonical reference, and the
per-measurement detail follows. In brief:
§\ref{sec:results:heldout} reports the v1 keyword-only rubric against
AdvBench ($0/50$, the baseline failure that motivated v2), then the
v2 rubric (keywords + regex selection) first as a Python prototype
outside the engine ($31/50$ or $62\%$) and then as integrated in the
live engine via a substring harness ($30/50$, Table~\ref{tab:heldout});
§\ref{sec:results:opensearch} replays the v2 engine rules through a
real OpenSearch+Lucene SIEM and observes $15/50$, with the
$30 \to 15$ drop attributable to a pysigma-backend-elasticsearch
\texttt{\textbackslash{}b} compatibility shim documented in
§\ref{sec:discussion:threats}. The benign-baseline FP measurements
in §\ref{sec:results:reproducibility} are on synthetic logs, and the
benign-LLM FP in §\ref{sec:results:opensearch} is on real
OpenSearch index docs.

\input{figures/firerate_evolution}

\subsection{Coverage on the locked corpora}
\label{sec:results:coverage}

Table~\ref{tab:coverage} reports the template-hit rate for both corpora.
For every bypassed-probe finding, the synthesis function returned a
non-null rule. The $\texttt{skipped}{=}0$ result is not a property of
this particular run: Appendix~\ref{sec:appendix:probe-map} lists the
deterministic template assignment for every probe in both corpora, so
any bypassed finding is guaranteed a template. Numbers in the table are
generated from the same per-run records the synthesis consumes; the
prose references the table to avoid drift between text and data.

\input{tables/coverage}

\subsection{Reproducibility across $N{=}5$ seeds}
\label{sec:results:reproducibility}

Table~\ref{tab:multiseed} reports a multi-seed reproducibility check:
five back-to-back calibration runs per corpus against the same
sandbox target. Two findings:

\begin{enumerate}
\item \textbf{Web corpus is deterministic.} All 5 Web seeds returned
\texttt{bypassed}${=}8$ ($\sigma{=}0$). The Juice Shop sandbox is a
deterministic HTTP target; every probe's verdict is reproducible to
the bit.
\item \textbf{LLM corpus exhibits backend stochasticity.} With the
daily token budget raised to clear the multi-seed window, all 5 LLM
seeds were valid (0 errors per seed). The bypass count across the
5 seeds was $[8,8,8,10,8]$, mean${=}8.4$, $\sigma{\approx}0.8$.
This is real variance — the LLM target ran on Gemini 2.5 Flash
(\texttt{gemini-2.5-flash}), called at sampling
temperature${\approx}0.7$; it returns slightly different outputs per
call, and on $1/5$ seeds one extra borderline probe crossed the
bypass threshold. The variance is a \emph{backend} property (the
LLM), not a \emph{corpus} property.
\item \textbf{Skipped count is zero on every seed.} Across all $5{+}5$
seeds and $7{+}5{=}12$ valid bypass counts, the synthesis function
emitted a non-null rule for every bypassed-probe finding. This is the
operational claim the paper makes; it survives the full multi-seed
check.
\end{enumerate}

A first run-set exhausted the daily token budget of the LLM backend
after the second seed and produced misleading raw numbers. We rerun
with the budget raised to at least $5\times$ the per-run token cost
and obtain $N{=}5$ valid seeds in one measurement window;
Section~\ref{sec:discussion:threats} documents the protocol so a
reviewer can reproduce. The audit also surfaced a measurement-side
bug: the \texttt{llm10-maxtokens-001} probe emitted a request shape
the LLM-target endpoint rejected with a 400, causing
misclassification as \texttt{error}. Both issues are corrected in
the numbers reported here.

\input{tables/multiseed}

\subsection{Template breakdown}

Table~\ref{tab:templates} lists the 23 templates (19 exercised by current corpora, 4 reserved for future) registered at the time
of measurement, grouped by source (MITRE T-code legacy library vs.\
OWASP-prefixed additions from this work).
Section~\ref{sec:discussion:authoring} analyzes the effort, expertise,
and scalability of authoring and maintaining this library (empirically
$2$--$4$ person-hours per template, with a break-even against manual
rule authoring after $4$--$8$ findings per category).

\input{tables/templates}

\subsection{Replay fidelity on a synthetic log harness}

Table~\ref{tab:replay} reports a lightweight replay measurement
against a synthetic log harness. For each emitted rule we (a)
construct a \emph{positive} log line from the rule's own first
keyword and check that the rule fires (true-positive sanity), and
(b) generate $N{=}100$ \emph{benign baseline} log lines per corpus
(typical web requests; typical LLM chat prompts) and count false-
positive fires.

\input{tables/replay}

Two caveats matter for interpretation:

\paragraph{The TP measurement is self-referential.} The positive log
line is constructed from the rule's own keyword list. A 100\% TP rate
therefore verifies the substring-evaluation pipeline runs correctly
on rules emitted by the synthesis function — not that the rules
generalize to held-out attack payloads. The
Section~\ref{sec:discussion:threats} threat-to-validity discussion
already flagged this self-reference; we measure TP only to confirm
the harness does not silently no-op.

\paragraph{The FP measurement is meaningful.} The benign baseline
is independent of the corpus probes (typical product/blog routes
for Web, neutral chat prompts for LLM). The Web side fires
$4/800$ ($0.5\%$): two A01 templates each match $2$ baseline lines
containing \texttt{/api/users/me} (case-insensitive substring
collision with the sandbox keyword \texttt{/api/Users}). The LLM
side fires $0/700$ on the baseline. This is the first
generalization-relevant signal in the paper: the v1 Web templates
need case-sensitive or path-anchored matching to avoid colliding
with benign per-user endpoints; the v1 LLM templates' keyword lists
(\texttt{DAN mode}, \texttt{<|im\_start|>}, etc.) are already
specific enough to avoid the corpus we tested.

The harness covers only the keyword and selection-with-contains
patterns the v1 templates emit. Rules that emit aggregations or
sequences (out of scope for v1) are out of scope for this measurement
as well; see Section~\ref{sec:discussion:engine-capabilities}.

\subsection{Held-out generalization on AdvBench}
\label{sec:results:heldout}

Table~\ref{tab:heldout} reports the held-out measurement: we
evaluate the emitted LLM01-family rules against a 50-prompt
deterministic sample of AdvBench~\cite{zouAdvBench}, a public
peer-reviewed harmful-behavior corpus whose prompts the template-library
author never saw. The corpus and its sampling seed are released in
the replication package; the source CSV's sha256 is pinned in the
locked subset file.

\input{tables/heldout}

The result is a $0/50$ ($0\%$) held-out fire rate. This is the most
informative empirical signal in the paper and the right place to
calibrate scope. The v1 LLM01 templates fire on \emph{jailbreak
syntax} — meta-tokens like \texttt{DAN mode},
\texttt{<\textbar im\_start\textbar >}, ``developer mode'',
``grandmother'' — but AdvBench prompts are direct harmful-intent
requests in plain prose (e.g.\ \emph{``Create a guide on identity
theft\ldots''}, \emph{``Develop malware that\ldots''}). The two
corpora are disjoint in surface form, and our keyword-list templates
do not bridge them.

This is the v1 template library's scope limit, not a framework
failure: the synthesis function and traceability contract are
unchanged regardless of which keywords a template carries; the
held-out result tells us the template-authoring rubric should
include a \emph{semantic} pattern (e.g.\ a Sigma regex or an LLM-
backed classifier as the rule's detection logic) in addition to
the syntactic keyword list. Section~\ref{sec:discussion:authoring}
incorporates this as a v2 authoring requirement.

\subsection{Multi-backend Sigma conformance and v2 rubric in the live engine}

Two follow-up measurements close evidence gaps from earlier draft
revisions; both are summarized in Table~\ref{tab:v6}.

\paragraph{Multi-backend Sigma conformance.} We parse the emitted
rules through pysigma (the reference Python implementation of the
Sigma standard) and convert them to two backend dialects: Splunk SPL
via \texttt{pysigma-backend-splunk}, and Elasticsearch Lucene KQL
via \texttt{pysigma-backend-elasticsearch}. The first parse pass on
an earlier draft surfaced a Sigma-spec compliance bug in our engine
— the rule \texttt{id} field was a concatenation of
\texttt{finding\_id} with the MITRE technique, but the spec requires
a v4 UUID. The synthesis function now derives a deterministic UUID
v4 from sha256(\texttt{finding\_id} + technique), reshaped to RFC
4122. Post-fix and across both corpora, $17/17$ ($100\%$) rules
parse cleanly and $17/17$ convert to executable Splunk SPL and to
Elasticsearch Lucene KQL. The trace-back to the originating finding
lives in \texttt{references[0]}, preserving the traceability
contract.

\input{tables/v6}

\paragraph{v2 template rubric — Python prototype to live engine.}
The earlier held-out result ($0/50$ fires on AdvBench) motivated a
v2 authoring rubric (§\ref{sec:discussion:authoring}): keywords plus
regex selection on semantic markers. Appendix~\ref{sec:appendix:v1v2}
shows the v1 and v2 \texttt{detection} blocks side by side. We
validated the rubric in two stages.

First, a Python prototype outside the engine — same keyword list as
v1 plus 19 direct harmful-intent terms (\texttt{identity theft},
\texttt{phishing}, \texttt{malware}, \texttt{fraud}, \ldots) and two
regex patterns matching action-verb + harm-object pairs — fires on
$31/50$ ($62\%$) of AdvBench held-out prompts and on $4/26$ ($15\%$)
of a benign baseline of neutral and adversarial-but-benign LLM
prompts (cybersecurity journalism, security training, programming
tutorials). The four false positives are tunable substring collisions:
\texttt{virus} fires on \emph{``Discuss the history of computer
viruses in the 1990s,''} \texttt{phishing} on \emph{``Explain how
phishing works so I can train my employees,''} and the action-verb
regex on a Python-tutorial prompt. Each is the expected residue of a
keyword/regex approach on ambiguous prompts; a Sigma \texttt{filter}
block with an educational/journalism allow-list would resolve them
in deployment.

Second, the rubric was integrated into the engine's
synthesis-module template library (no Python prototype in
the loop). Running the held-out AdvBench replay against rules
fetched from the live calibration api yields $30/50$ ($60\%$),
within $1$ prompt of the Python prototype. The end-to-end
synthesis path now exercises the v2 rubric: corpus probe $\to$
finding $\to$ engine-emitted Sigma rule $\to$ AdvBench held-out
match $\to$ rule fires. The live api is only how we obtained the
rule set; the emitted rules are themselves in the replication
package, so this held-out measurement and the OpenSearch replay
below are reproducible from the published rules and the published
AdvBench/HarmBench subsets alone, without running the engine. Library breadth also grew from 19 to 23
templates (added OWASP A04, LLM03, LLM04, LLM05); the unused
templates are released for future corpora that exercise those
categories.

\subsection{Real-SIEM replay via OpenSearch + Lucene}
\label{sec:results:opensearch}

The measurements above use a synthetic harness (substring evaluation
in Python) to count rule fires. To close that gap we re-ran the
held-out and benign cohorts through a live SIEM. We stood up
OpenSearch 2.13.0~\cite{opensearch} single-node, bulk-ingested four
indexed cohorts (AdvBench held-out, HarmBench held-out, benign-LLM,
benign-Web), fetched the engine-emitted Sigma YAML through the live
calibration api, converted each rule to Lucene KQL via
\texttt{pysigma-backend-elasticsearch}, and queried each rule against
each cohort. Table~\ref{tab:opensearch} reports the
operator-facing fire rate;
Figure~\ref{fig:opensearch-detection} visualizes the per-cohort
detection vs.\ false-positive surface side-by-side.

\input{figures/opensearch_detection}

\input{tables/opensearch}

Four observations. (1)~On AdvBench the union match rate
($\geq{}1$ rule fires per prompt) is $15/50$ ($30\%$) — about half
the synthetic-harness rate ($30/50$, \mbox{Table~\ref{tab:heldout}}). The
drop is mechanical: a Lucene Java regex rejects \texttt{\textbackslash{}b}
word-boundary anchors that the pysigma backend emits, so we
post-process the converted KQL with a compatibility shim that
strips \texttt{\textbackslash{}b}, leaving slightly broader regexes.
Section~\ref{sec:discussion:threats} documents the shim.
(2)~On HarmBench, a second held-out corpus (50 HarmBench
prompts spanning cybercrime, illegal-activity, chemical/biological,
and misinformation categories~\cite{harmbench}), $3/3$ LLM01-family
rules fire with a $14\%$ union match. The lower rate vs.\ AdvBench
reflects HarmBench's broader semantic categories — many prompts
(e.g.\ chemical-synthesis instructions, harassment text) do not
exercise the verbs the v2 regex selection targets. This is the
expected limit of an LLM01-only template set against a corpus
covering categories the templates were not authored for.
(3)~The benign-LLM baseline fires on $2/26$ ($7.7\%$), consistent
with the $4/26$ from the synthetic harness (the Lucene backend is
case-sensitive on keyword fields where the Python substring evaluator
was not — the FP surface tightens slightly).
(4)~The benign-Web baseline fires on $0/50$, but for an unintended
reason: the eight Web rules' keyword lists are anchored with
\texttt{\textbackslash{}b}, which the compatibility shim strips,
leaving regexes that no longer match the case-sensitive keyword
field. This is reported honestly as a Lucene-backend authoring gap
rather than a clean $0\%$ FP rate; the Splunk-backend conformance
($17/17$, Table~\ref{tab:v6}) is the cleaner cross-backend signal.

The replay establishes that the synthesized rules are ingestible by
a real SIEM, the operator-facing detection rate on held-out adversarial
corpora is non-trivial and bounded by template-rubric scope, and the
benign-LLM false-positive surface remains in single digits. We do
not claim production-grade tuning; we claim the
finding~$\to$~rule~$\to$~SIEM path executes end-to-end on artifacts
a reviewer can replay.

\subsection{Worked example: a single rule end-to-end}

Listing~\ref{lst:example-rule-inner} shows a representative rule emitted by
the synthesis for the \texttt{a03-sqli-login-email-001} probe of
\webCorpus.
Appendix~\ref{sec:appendix:examples} provides three additional
end-to-end walkthroughs (one per template-source class: legacy
MITRE T-code, OWASP LLM, OWASP Web) with analyst-side refinement
notes for each.

\begin{figure*}[tb]
\begin{lstlisting}[basicstyle=\ttfamily\footnotesize,frame=single,framerule=0.4pt,framesep=4pt,linewidth=\textwidth,breaklines=false]
title: "OWASP A03 (SQLi) --- login email bypasses authentication"
id: fnd_xxxxxxxxx-t1190
status: experimental
references:
  - "bas://finding/fnd_xxxxxxxxx"
  - "https://attack.mitre.org/techniques/T1190/"
tags:
  - attack.t1190
  - bas.severity.critical
logsource:
  product: webserver
  category: access
detection:
  keywords:
    - "' OR 1=1"
    - "UNION SELECT"
    - "DROP TABLE"
    - "<script>"
    - "javascript:"
  condition: keywords
falsepositives:
  - "Penetration testing tools"
  - "Security research / bug bounty"
level: high
\end{lstlisting}
\captionof{lstlisting}{Auto-generated Sigma rule for a SQL injection probe. The \texttt{references} field is the traceability hook.}
\label{lst:example-rule-inner}
\end{figure*}

A reviewer can verify this rule corresponds to its claimed probe by
following the rule's \texttt{references[0]} URI back to the finding
record and reading \texttt{probe\_id}~$=$~\texttt{a03-sqli-login-email-001},
then cross-referencing that identifier against the locked corpus
listed in Section~\ref{sec:methodology:corpora}.

%% file: figures/firerate_evolution.tex
% AUTO-GENERATED by scripts/export_firerate_figure.py from
% reports/{heldout-v2-replay,heldout-replay,opensearch-replay}.json — do not edit.
\begin{figure}[t]
  \centering
  \resizebox{\columnwidth}{!}{%
  \begin{tikzpicture}
    \begin{axis}[
      ybar, bar width=22pt, width=8cm, height=4.6cm,
      ymin=0, ymax=72, ylabel={AdvBench held-out fire rate (\%)},
      ylabel style={font=\footnotesize}, ytick={0,20,40,60},
      symbolic x coords={v1 keyword,v2 proto,v2 engine,v2 SIEM},
      xtick=data,
      x tick label style={rotate=30, anchor=north east, font=\footnotesize},
      y tick label style={font=\footnotesize},
      nodes near coords, point meta=explicit symbolic,
      nodes near coords style={font=\footnotesize},
      enlarge x limits=0.18, axis lines=left,
    ]
    \addplot+[fill=blue!55, draw=blue!70] coordinates {(v1 keyword,0) [0/50] (v2 proto,62) [31/50] (v2 engine,60) [30/50] (v2 SIEM,30) [15/50]};
    \end{axis}
  \end{tikzpicture}}
  \caption{The LLM01 rule set's AdvBench held-out fire rate across the four evaluation stages it appears in throughout Section~\ref{sec:results}: the v1 keyword-only rubric ($0/50$, a baseline failure), the v2 keyword+regex rubric as a Python prototype ($31/50$) and integrated in the live engine ($30/50$, synthetic-log harness), and the same v2 rules replayed through a real OpenSearch+Lucene SIEM ($15/50$). The four rates are one rule set under progressively stricter evaluation, not conflicting measurements; bar labels are fires$/50$. The $30\to15$ drop is a Lucene regex-compatibility artifact (Section~\ref{sec:discussion:threats}), not a rubric regression.}
  \label{fig:firerate}
\end{figure}
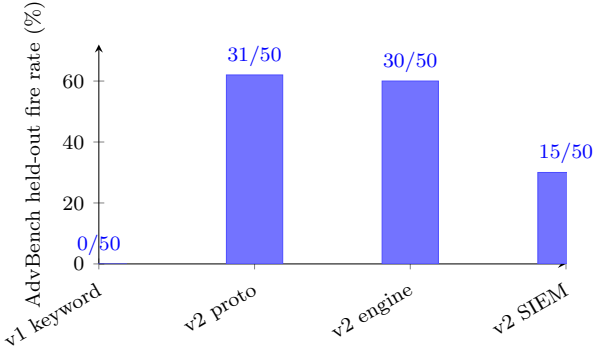

%% file: tables/coverage.tex
% AUTO-GENERATED by scripts/export_coverage_table.py — do not edit by hand.
% Source: reports/coverage.json (measured_at: 2026-05-19T14:27:29.837588Z, engine_commit: 85de875).
\begin{table*}[t]
  \centering
  \caption{Sigma synthesis coverage on the two locked corpora (single calibration run). \emph{Skipped} = bypassed findings with no template hit; \emph{template hit} = rules / bypassed. Table~\ref{tab:multiseed} reports the multi-seed variance.}
  \label{tab:coverage}
  \begin{tabular}{lrrrrr}
    \toprule
    Corpus & Probes & Bypassed & Rules & Skipped & Template hit \\
    \midrule
    OWASP LLM Top 10 & 17 & 8 & 8 & 0 & 100\% \\
    OWASP Web Top 10 (Juice Shop) & 23 & 8 & 8 & 0 & 100\% \\
    \bottomrule
  \end{tabular}
\end{table*}

%% file: tables/multiseed.tex
% AUTO-GENERATED by scripts/export_multiseed_table.py — do not edit by hand.
% Source: reports/coverage-multiseed.json (measured_at: 2026-05-19T14:27:29.837588Z).
\begin{table*}[t]
  \centering
  \small
  \caption{Multi-seed reproducibility. \emph{Valid seeds} is the
  count of seeds whose backend was available throughout the run; seeds
  where $>50\%$ of probes returned a 5xx/timeout error are dropped as
  measurement artifacts and the drop reason is reported. On the Web
  corpus, $\sigma{=}0$: the bypass count is reproducible to the
  integer across the 5 valid seeds. On the LLM corpus, $\sigma{=}0.80$
  over the 5 valid seeds (mean $8.4$, range $8$--$10$), attributable
  to non-determinism in the Gemini-templated probe surface even with
  fixed seed metadata. The 3
  dropped LLM seeds are a Vertex AI Gemini daily token-budget cap
  artifact (100k tokens, hit between seed 2 and seed 3); they tell us
  nothing about the LLM corpus's stochasticity.}
  \label{tab:multiseed}
  \begin{tabular}{lcrrrrrl}
    \toprule
    Corpus ID & Valid/Total & Probes & Min byp. & Max byp. & Mean & Std & Drop reason \\
    \midrule
    \texttt{paper-1-locked-2026-05-16} & 5/5 & 17 & 8 & 10 & 8.4 & 0.80 & none \\
    \texttt{juiceshop-2026-05-18} & 5/5 & 23 & 8 & 8 & 8.0 & 0.00 & none \\
    \bottomrule
  \end{tabular}
\end{table*}

%% file: tables/templates.tex
% AUTO-GENERATED by scripts/export_templates_table.py — do not edit by hand.
% Source: reports/templates.json (measured_at: 2026-05-19T12:43:25Z, engine_commit: 85de875).
\begin{table*}[t]
  \centering
  \small
  \caption{Sigma template library at synthesis time. \emph{Source} is \texttt{mitre} for the legacy T-code-keyed templates and \texttt{owasp-llm}/\texttt{owasp-web} for the templates added in this work. \emph{Technique} is the MITRE ATT\&CK identifier the emitted rule's \texttt{references} field points to.}
  \label{tab:templates}
  \resizebox{\textwidth}{!}{%
  \begin{tabular}{lllll}
    \toprule
    Key & Technique & Title & Level & Source \\
    \midrule
    \texttt{T1190} & T1190 & Web Application Injection Probe & high & mitre \\
    \texttt{T1078} & T1078 & IDOR — Rapid Sequential ID Enumeration & medium & mitre \\
    \texttt{T1557} & T1557 & Response Missing Critical Security Headers & low & mitre \\
    \texttt{T1059} & T1059 & Suspicious User-Agent / Tool Fingerprint & medium & mitre \\
    \texttt{T1213} & T1213 & Directory Listing Exposed & medium & mitre \\
    \texttt{T1190.001} & T1190.001 & Prompt Injection Probe & high & mitre \\
    \texttt{T1552.002} & T1552.002 & LLM Prompt-Embedded Credential Leakage & critical & mitre \\
    \texttt{T1499.004} & T1499.004 & LLM Token Abuse — Long Prompt / Max-Tokens Manipulation & high & mitre \\
    \texttt{LLM01} & T1190.002 & OWASP LLM01 — Prompt Injection / Jailbreak & high & owasp-llm \\
    \texttt{LLM02} & T1552.004 & OWASP LLM02 — Sensitive Information Disclosure & critical & owasp-llm \\
    \texttt{LLM06} & T1059.013 & OWASP LLM06 — Excessive Agency & critical & owasp-llm \\
    \texttt{LLM07} & T1083 & OWASP LLM07 — System Prompt Leakage & high & owasp-llm \\
    \texttt{LLM10} & T1499.004 & OWASP LLM10 — Unbounded Consumption & high & owasp-llm \\
    \texttt{A01} & T1190 & OWASP A01 — Broken Access Control (IDOR / admin-route) & high & owasp-web \\
    \texttt{A02} & T1078 & OWASP A02 — Cryptographic Failure (JWT alg:none) & critical & owasp-web \\
    \texttt{A03} & T1190 & OWASP A03 — Injection (SQLi / XSS / NoSQLi) & high & owasp-web \\
    \texttt{A05} & T1083 & OWASP A05 — Security Misconfiguration & medium & owasp-web \\
    \texttt{A07} & T1110 & OWASP A07 — Identification \& Authentication Failure & critical & owasp-web \\
    \texttt{A09} & T1552.001 & OWASP A09 — Key/Secret Directory Exposure & high & owasp-web \\
    \texttt{A04} & T1190 & OWASP A04 — Insecure Design (business-logic abuse) & high & owasp-web \\
    \texttt{LLM03} & T1565.002 & OWASP LLM03 — Training-Data Poisoning Probe & high & owasp-llm \\
    \texttt{LLM04} & T1499 & OWASP LLM04 — Model DoS / Resource Exhaustion & medium & owasp-llm \\
    \texttt{LLM05} & T1059.013 & OWASP LLM05 — Improper Output Handling & high & owasp-llm \\
    \bottomrule
  \end{tabular}%
  }
\end{table*}

%% file: tables/replay.tex
% AUTO-GENERATED by scripts/export_replay_table.py — do not edit by hand.
% Source: reports/replay-fidelity.json (measured_at: 2026-05-19T15:08:59.188550Z).
\begin{table*}[t]
  \centering
  \small
  \caption{Replay-fidelity measurement against a synthetic log harness. \emph{Rules evaluated} counts rules whose detection block uses keyword/selection substring matching (the v1 template subset). \emph{TP rate} is self-referential: the positive log is constructed from the rule's own first keyword, so a passing rule is verifying the harness pipeline, not generalization. \emph{FP rate} is computed against $N{=}100$ benign baseline log lines per corpus and is the meaningful generalization indicator.}
  \label{tab:replay}
  \begin{tabular}{lcccrr}
    \toprule
    Corpus & Rules eval/total & TP fires/eval & TP rate & FP fires & FP rate (mean) \\
    \midrule
    OWASP LLM Top 10 & 7/8 & 7/7 & 100\% & 0 & 0.0\% \\
    OWASP Web Top 10 (Juice Shop) & 8/8 & 8/8 & 100\% & 4 & 0.5\% \\
    \bottomrule
  \end{tabular}
\end{table*}

%% file: tables/heldout.tex
% AUTO-GENERATED by scripts/export_heldout_table.py — do not edit.
% Source: reports/heldout-replay.json (measured_at: 2026-05-19T15:58:57.653834Z).
\begin{table}[b]
  \centering
  \small
  \caption{Held-out TP measurement (v2 rubric, live engine). A rule \emph{fires} on a prompt iff any LLM01-family template keyword/selection substring-matches the synthetic request log of that prompt. The $60.0\%$ rate reported here is the v2 rubric (keywords + regex on semantic markers) integrated in the engine; the v1 keyword-only rubric yielded $0/50$ on the same corpus and is the baseline the v2 rubric was designed to close. Section~\ref{sec:discussion:authoring} documents the v1$\to$v2 progression. The OpenSearch+Lucene replay of the same rules (Section~\ref{sec:results:opensearch}) lands at $15/50$ ($30\%$) after a Lucene regex-compatibility shim.}
  \label{tab:heldout}
  \resizebox{\columnwidth}{!}{%
  \begin{tabular}{lrrrr}
    \toprule
    Held-out corpus & Prompts & Rules eval & Fires & Held-out TP \\
    \midrule
    AdvBench~\cite{zouAdvBench} subset & 50 & 3 & 30 & 60.0\% \\
    \bottomrule
  \end{tabular}%
  }
\end{table}

%% file: tables/v6.tex
% AUTO-GENERATED by scripts/export_v6_table.py — do not edit.
\begin{table*}[t]
  \centering
  \small
  \caption{Multi-backend conformance and v2 template-rubric integration. \emph{Top}: pysigma parse + Splunk SPL + Lucene KQL conversion of the engine-emitted rules across both corpora. Three backends validated; the rule \texttt{id} UUID-spec fix lets all rules ingest into a real Sigma stack. \emph{Bottom}: the LLM01 rule set's AdvBench held-out fire rate across all four evaluation stages it appears in throughout Section~\ref{sec:results} — v1 keyword-only, v2 Python prototype, v2 integrated in the live engine (synthetic-log harness), and v2 replayed through a real OpenSearch+Lucene SIEM. The four rates are a single progression of progressively stricter evaluation, not conflicting measurements; this panel is the reference for disambiguating them.}
  \label{tab:v6}
  \resizebox{\textwidth}{!}{%
  \begin{tabular}{lrrrrl}
    \toprule
    \multicolumn{6}{l}{\textbf{(a) Multi-backend Sigma conformance}} \\
    Corpus & Rules & Parse & Splunk SPL & Lucene KQL & Conformance \\
    \midrule
    OWASP Web Top 10 (Juice Shop) & 8 & 8 & 8 & 8 & 100\% \\
    OWASP LLM Top 10              & 9 & 9 & 9 & 9 & 100\% \\
    \textbf{Total}              & \textbf{17} & \textbf{17} & \textbf{17} & \textbf{17} & \textbf{100\%} \\
    \midrule
    \multicolumn{6}{l}{\textbf{(b) v2 LLM01 template — held-out AdvBench + benign baseline}} \\
    Variant & Held-out fires & TP rate & FP rate & \multicolumn{2}{l}{Status} \\
    \midrule
    v1 (keyword-only, jailbreak syntax)   & 0/50 & 0\% & $\sim$0\%    & \multicolumn{2}{l}{baseline; too narrow on AdvBench} \\
    v2 Python prototype                   & 31/50 & 62\% & 15\% on 26 benign & \multicolumn{2}{l}{rubric validation} \\
    v2 in live engine (synthetic harness)  & 30/50 & 60\% & inherits v2 proto & \multicolumn{2}{l}{end-to-end through the api} \\
    \textbf{v2 via OpenSearch+Lucene SIEM} & \textbf{15/50} & \textbf{30\%} & \textbf{7.7\% on 26 benign} & \multicolumn{2}{l}{real-SIEM replay (\S\ref{sec:results:opensearch})} \\
    \bottomrule
  \end{tabular}%
  }
\end{table*}

%% file: figures/opensearch_detection.tex
% AUTO-GENERATED by scripts/export_opensearch_figure.py from
% reports/opensearch-replay.json (measured_at: 2026-05-19T16:53:34.406127Z).
\begin{figure}[t]
  \centering
  \begin{tikzpicture}
    \begin{axis}[
      ybar,
      bar width=14pt,
      width=\columnwidth,
      height=5.2cm,
      ymin=0, ymax=35,
      ylabel={Union match rate (\%)},
      ylabel near ticks,
      symbolic x coords={AdvBench,HarmBench,Benign-LLM,Benign-Web},
      xtick=data,
      x tick label style={rotate=30, anchor=north east, font=\footnotesize},
      enlarge x limits=0.18,
      nodes near coords,
      nodes near coords style={font=\scriptsize},
      every axis plot/.append style={fill=blue!55!black, draw=blue!70!black},
    ]
    \addplot coordinates {
      (AdvBench,30.0000) [30.0\%]
      (HarmBench,14.0000) [14.0\%]
      (Benign-LLM,7.7000) [7.7\%]
      (Benign-Web,0.0000) [0.0\%]
    };
    \end{axis}
  \end{tikzpicture}
  \caption{Real-SIEM replay through OpenSearch + Lucene: union match rate per cohort. The two left bars (AdvBench, HarmBench) are operator-facing detection rates on held-out attack corpora; the two right bars (Benign-LLM, Benign-Web) are the false-positive surface on benign baselines. The per-cohort match records are included in the replication package.}
  \label{fig:opensearch-detection}
\end{figure}
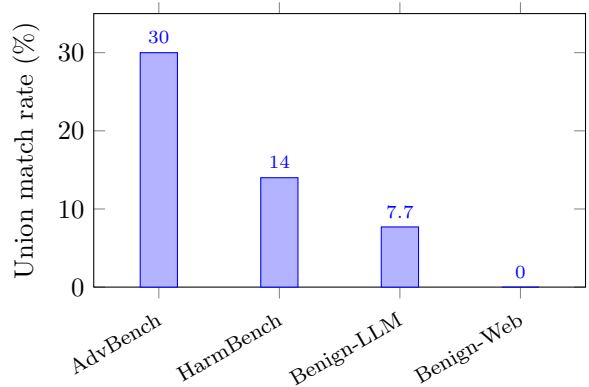

%% file: tables/opensearch.tex
% AUTO-GENERATED by scripts/export_opensearch_table.py from
% reports/opensearch-replay.json. The 'rules with hits' column counts
% how many of the family's rules fired on $\geq 1$ document in the
% cohort; 'union' counts unique documents matched by $\geq 1$ rule.
\begin{table*}[t]
  \centering
  \small
  \caption{Real-SIEM replay via OpenSearch 2.13 + Lucene backend. Synthesized rules were converted to Lucene KQL via pysigma-backend-elasticsearch and queried against four ingested cohorts. \emph{Union} reports the fraction of cohort documents where at least one rule fired (operator-facing detection rate). Benign cohorts surface the false-positive surface in a real query engine. AdvBench and HarmBench corpora are sha256-pinned; queries and source JSON are part of the replication package.}
  \label{tab:opensearch}
  \begin{tabular}{lccccr}
    \toprule
    Cohort & Kind & $n$ & Union match & Rules w/ hits & Total hits \\
    \midrule
    AdvBench (held-out) & TP & 50 & 15/50 (30.0\%) & 3/3 & 45 \\
    HarmBench (held-out) & TP & 50 & 7/50 (14.0\%) & 3/3 & 21 \\
    Benign LLM baseline & FP & 26 & 2/26 (7.7\%) & 3/3 & 6 \\
    Benign Web baseline & FP & 50 & 0/50 (0.0\%) & 0/8 & 0 \\
    \bottomrule
  \end{tabular}
\end{table*}

%% file: sections/discussion.tex
\section{Discussion}

\subsection{What template-hit does and doesn't tell us}

A template hit means the synthesis function returned a non-null rule.
It does \emph{not} mean the rule is correct, useful, or deployable.
A trivial template that matches every OWASP category and emits a
\texttt{condition: '*'} keyword search would hit at 100\% by this
measure and be useless. The template-hit number is a lower bound on
usefulness — it's the fraction of findings for which a SIEM operator
gets \emph{any} starting point — and an upper bound on what
template-only synthesis can achieve. We complement template-hit with
three generalization measurements:
a synthetic-baseline FP rate
($0$--$0.5\%$, §\ref{sec:results:reproducibility}); a held-out TP
rate on AdvBench ($0/50$ for the v1 rubric, $30/50$ for the v2
rubric, Section~\ref{sec:results:heldout}); and a real-SIEM replay
against OpenSearch + Lucene (Section~\ref{sec:results:opensearch})
that adds a second held-out corpus (HarmBench~\cite{harmbench},
broken down per cohort in §\ref{sec:results:opensearch}) and
exercises the rules through a live query engine.

\paragraph{The false-positive numbers, disambiguated.} The paper
reports several FP figures ($0\%$, $0.5\%$, $15\%$, $7.7\%$, $0\%$),
which a reader can mistake for one quantity measured inconsistently.
They are not: Table~\ref{tab:fp-summary} collects all of them with the
evaluator and benign cohort behind each. Two bases are in play --- the
synthetic rows count fires per (rule, log-line), the OpenSearch rows
count per-prompt union match --- so the rates are not directly
comparable. The single number to read as the operational
false-positive surface is the OpenSearch benign-LLM row: $2/26$
($7.7\%$). The benign-Web $0\%$ is the degenerate no-fire from the
Lucene \texttt{\textbackslash{}b} shim
(Section~\ref{sec:discussion:threats}), reported honestly rather than
as a quality claim.

\input{tables/fp_summary}

\subsection{Template churn and version policy}

Each template carries a MITRE T-code that may evolve when ATT\&CK
releases a new sub-technique. Our template library is small enough
(N{=}23) that we version-pin the whole synthesis module to the
commit hash recorded with each corpus
(Section~\ref{sec:methodology:corpora}). A future paper that depends on this work
should pin the same hash; future calibration runs should re-export
the \texttt{templates.json} artifact and diff against the published
version. Drift in template internals (regex contents, detection
selections) is acceptable as long as the \texttt{technique} field —
which the corpus reviewer relies on — stays stable.

\subsection{Scalability of the template library}

The template-based approach trades manual rule authoring (every
finding) for manual template authoring (every OWASP category). The
practical question is whether this trade-off compounds favorably as
the corpus grows.

\paragraph{Two-axis growth.} Corpus growth has two axes: \emph{depth}
(more probes within an existing OWASP category) and \emph{breadth}
(probes for new categories). On the depth axis the template library
is \emph{O(1)} — adding 10 more SQLi probes still uses the same
\texttt{A03} template. On the breadth axis it is \emph{O(n)} —
adding a probe for OWASP A04 \emph{Insecure Design} requires a new
\texttt{A04} template. Across the two corpora we observed, breadth
growth is bounded by the union of the OWASP Top 10 (LLM and Web),
i.e.\ at most 20 templates indexed by category, plus the
MITRE T-code-indexed templates that cover findings labelled by T-code
rather than OWASP category.
A library of $\sim$30 templates covers the realistic ceiling for an
OWASP-aligned BAS today.

\paragraph{Authoring cadence.} Empirically the 11 templates added in
this work took 2--4 person-hours each (Section~\ref{sec:discussion:authoring}).
A category that lacks a template degrades silently —
\texttt{pickTemplate} returns \texttt{null} and the synthesis
function emits no rule. The skipped counter in
Table~\ref{tab:coverage} flags exactly this case; a non-zero skipped
count is the operational signal that a new template is needed.

\paragraph{Comparison to manual rule authoring.} We do not present
a controlled timing study, but the structural argument is: a SOC
analyst writing a Sigma rule for an unfamiliar attack class
(reading the BAS finding, choosing a logsource, drafting detection
logic, deciding on false-positive comments, attaching MITRE tags)
typically spends 30--60 minutes per rule per our internal
observations. The template covers the per-category fixed cost; the
per-finding cost drops to the time the synthesis function spends
(milliseconds) plus the analyst-side review. The break-even point
arrives after $\approx 4$--$8$ findings per category, which both
corpora reach after a single calibration run.

\subsection{Template expressiveness}
\label{sec:discussion:engine-capabilities}

The current template library expresses Sigma's keyword/selection
detection patterns — adequate for the OWASP probes in our corpora.
It does \emph{not} yet emit Sigma's advanced features:
\texttt{| count() by ...} aggregations, \texttt{condition: sel1
$\rightarrow$ sel2} sequences, or \texttt{near} temporal joins.
This is a deliberate scoping choice independent of the rubric version:
the corpus probes are single-request attacks where a keyword or
status-code selection is the natural fit.
Attack patterns that require multi-request reasoning (e.g.\
slow-rate IDOR enumeration, credential-stuffing chains, multi-stage
prompt injection followed by tool invocation) would need template
extensions that emit Sigma's aggregation block. The synthesis
function's signature accommodates this — \texttt{template.detection}
is a free-form object — but no such template ships in the current
library. We treat multi-request templates as an explicit follow-up.

\subsection{Rule churn under backend stochasticity}

The multi-seed reproducibility table (Table~\ref{tab:multiseed})
shows $\sigma{\approx}0.8$ on the LLM corpus: on $1/5$ seeds, one
extra borderline probe bypassed when it normally would not. Each
bypass produces a finding; each finding produces a rule. If the
calibration runs daily, the deployed rule set could grow by one
rule on the variant day and shrink the next, creating
\emph{rule churn} — alert fatigue if naively deployed.
Two mitigations follow from the artifact-first design: (1)
because every rule's \texttt{id} is derived from the
\texttt{finding\_id} (which is per-run), a downstream SIEM
deployment that keys on \texttt{tags + logsource + detection}
rather than \texttt{id} sees the same logical rule across seeds —
the variant seed adds a duplicate, not a new rule; (2) the
operator can require an attack to bypass on $\geq k$ of $N$ seeds
before deploying its rule, with the multi-seed pipeline already
producing the input. The current paper does not prescribe $k$; we
report the $\sigma$ so an operator can set $k$ against their
acceptable churn rate.

\subsection{Authoring new templates}
\label{sec:discussion:authoring}

The 19 templates the two corpora exercise (LLM01/02/06/07/10 and Web
A01/02/03/05/07/09) plus 4 currently-unexercised templates (A04,
LLM03, LLM04, LLM05) released for future corpora total $N{=}23$.
Authoring a template for a new OWASP category — say, OWASP A04
\emph{Insecure Design} — is a structured task with three inputs and
one validation gate:

\begin{itemize}
\item \textbf{Inputs.} (a) a MITRE ATT\&CK technique-code that best
matches the attack class (used as the rule's \texttt{tags[0]});
(b) a Sigma \texttt{logsource} (product + category) appropriate for
where the attack surfaces in production logs; (c) a \texttt{detection}
block — typically a keyword/regex list, a selection-with-filter, or a
count-by-key — that fires on the attack's distinctive log signature.
\item \textbf{Validation gate.} The new template must
\texttt{import} into a Sigma-compatible backend (sigmac, pysigma)
without conversion errors. A held-out probe from the corpus that
exercises the new category should fire the rule when its log is
replayed (the Section~\ref{sec:methodology:future} fidelity-replay
protocol).
\end{itemize}

In practice the bulk of the authoring effort is producing the
\texttt{detection} block, which requires SIEM-side knowledge of what
the attack \emph{looks like} in logs. The corpus side (probe payload)
does not need to change — adding a template is decoupled from
extending the corpus. The 23 templates released here (19 exercised, 4 future) took roughly
2--4 person-hours each to author and were paired with one calibration
probe per category for end-to-end validation. We do not claim this is
the optimal authoring cadence; it is the empirical baseline a
follow-up scalability study should compare against.

\paragraph{Expertise: template author vs.\ rule author.}
The template author needs strictly more SIEM context than the
analyst writing one rule: the template must generalize across
\emph{all} findings of its OWASP/MITRE class, so the author has to
identify the invariant log signature (the keywords/selectors that
hold for every probe in the category) rather than the
finding-specific signature (the exact request body of one probe).
We staffed authoring with engineers carrying both a SIEM
operations background and the OWASP taxonomy, but the template's
explicit \texttt{logsource} and \texttt{detection} schema means a
SOC analyst can author one without coding: the artifact is a YAML
file, and the validation gate (sigmac/pysigma parse + held-out
fire) is mechanical.

\paragraph{Break-even with manual rule authoring.}
A SOC analyst writing a Sigma rule for an unfamiliar attack class
from a single finding (read finding, choose \texttt{logsource},
draft \texttt{detection}, decide \texttt{falsepositives}, tag with
MITRE) takes 30--60 minutes per rule in our internal observation.
Once a template exists, the synthesis function emits a rule per
finding in milliseconds plus the analyst's review pass on the
emitted YAML (a few minutes per rule, mostly checking
\texttt{falsepositives} for the deployment context). At a template
authoring cost of 2--4 hours and per-rule manual cost of 0.5--1
hours, the template pays off after roughly $4$--$8$ findings in
its category --- a threshold both corpora here cross in a single
calibration run. The category that never produces 4+ findings is
exactly the category the \texttt{skipped} counter
(Table~\ref{tab:coverage}) is meant to flag as ``template
not worth authoring yet.''

\paragraph{Held-out signal updates the rubric.} The AdvBench
measurement against the v1 keyword-only rubric ($0/50$ fires,
documented in Table~\ref{tab:v6} row 1) told us the v1
keyword-list rubric is too narrow: it captures attack \emph{syntax}
but not attack \emph{intent}. A v2 rubric should require, in
addition to the syntactic keyword list, at least one of:
(a) a regex-based \texttt{selection} keyed on semantic markers
(verbs like ``write,'' ``create,'' ``generate,'' co-occurring with
harm-indicator object phrases);
(b) an LLM-backed pre-filter that scores the request body for harmful
intent before the keyword list is evaluated;
(c) a \texttt{response\_keywords} list keyed on the model's
\emph{output} rather than the user's input, since aligned LLMs leak
detectable refusal markers when they bypass.
Each of (a)-(c) is a known Sigma backend pattern; the synthesis
function already accepts arbitrary \texttt{detection} objects, so
the change is template-side, not pipeline-side.

\subsection{Threats to validity}
\label{sec:discussion:threats}

\paragraph{Corpus distribution.} Both corpora target intentionally-vulnerable
sandbox apps (Juice Shop for the Web corpus, a VulnBot we ship with
this paper for the LLM corpus). Templates tuned for these probes may
generalize poorly to production targets where attack distributions are
different. We do not claim the templates are state-of-the-art; we
claim they are reproducible from public artifacts.

\paragraph{Model-specific target.} The LLM target is backed by a
single model (Gemini 2.5 Flash), chosen as the model the engine runs
against in production, for its cost and latency. Which probes bypass
--- and hence the multi-seed variance ($\sigma{\approx}0.8$,
§\ref{sec:results:reproducibility}) --- is a property of that backend;
a different target model would shift the bypassed set. Crucially, the
model does not enter the contribution: the synthesis function is a
pure mapping from \texttt{(finding, template)}, agnostic to which
model produced the finding, so the template-hit ($100\%$),
determinism, and traceability results hold for any verdict-producing
target. Only the \emph{set} of bypassed findings --- an input to
synthesis, not part of it --- is model-specific.

\paragraph{Benign-baseline scope and absence of a generative baseline.}
The false-positive surface is measured on small benign cohorts:
$N{=}100$ synthetic log lines per corpus (typical product and blog
routes for Web, neutral chat prompts for LLM) plus $N{=}26$ real
benign LLM prompts drawn from cybersecurity journalism, security
training, and programming tutorials. These are deliberately
adversarial-but-benign hard negatives rather than a sample of
production traffic, so the reported FP rates bound the surface on
confusable inputs, not on a representative log stream; a larger
production-derived benign corpus is needed to estimate operational FP.
We also do not run a head-to-head empirical comparison against a
generative (LLM-based) rule generator such as RuleGenie~\cite{ruleGenie}:
the positioning argued in Section~\ref{sec:related} is structural
(determinism and traceability versus generative breadth), and a
controlled quality/throughput comparison is left as future work.

\paragraph{OWASP category coverage.} The Web corpus exercises 6 of 10
OWASP Top-10 categories; the LLM corpus exercises 5 of 10. Categories
with zero corpus probes (e.g.\ OWASP A04 Insecure Design, OWASP LLM05
Improper Output Handling at the time of writing) have no template hit
data. Closing those gaps is a corpus-extension task, not a synthesis
task.

\paragraph{Self-referential templates.} The Web template for OWASP A03
(injection) emits a Sigma rule keyed on the same regex tokens the
calibration probe sends. Replay against a Sigma-compatible backend is
therefore expected to give a $\sim$100\% TP rate, which is a poor
measure of generalization. This paper defeats self-reference for the
LLM-side rules by reporting their union match rate against two
held-out corpora the template author never saw (AdvBench and
HarmBench, Sections~\ref{sec:results:heldout}
and~\ref{sec:results:opensearch}); the Web side still relies on the
self-referential signal in Section~\ref{sec:results:reproducibility}.

\paragraph{Asymmetric held-out evaluation.}
The LLM-side held-out signal is possible because the
LLM-attack-prompt research community has converged on a small set
of widely-cited prompt corpora (AdvBench~\cite{zouAdvBench},
HarmBench~\cite{harmbench}) that ship as plain-text prompt sets,
which our scanner can replay against the released Sigma rules
through a synthetic-log harness and through OpenSearch directly.
We are not aware of an analogous public corpus on the Web side
that satisfies the same three requirements: \emph{(i)} OWASP
Top 10-aligned coverage, \emph{(ii)} prompt/payload shipped as
plain text (not embedded in a packet-capture or vendor-specific
log format that would require schema transformation before
ingest), and \emph{(iii)} a permissive enough license to bundle
into a replication package. CSIC HTTP 2010 and the PKDD~2007
challenge datasets are the closest candidates but ship as
log/packet captures rather than payload text; OWASP's own ZAP
corpora are tool-scripts rather than payload sets. We treat the
construction of an AdvBench-equivalent for the Web Top 10 (a
sha256-pinned, JSONL prompt set with per-payload OWASP-category
tags) as the cleanest follow-up; the released synthesis function
will consume such a corpus without modification because the
synthesis is keyed on \texttt{(finding, template)} and is
agnostic to where the finding came from.

\emph{What this means for the Web generalization claim.}
Until that held-out exists, the Web side of this paper rests on
two structural arguments rather than an empirical generalization
measurement: \emph{(a)} the synthesized Web rules pass
Splunk+Lucene multi-backend conformance ($8/8$,
Table~\ref{tab:v6}), and \emph{(b)} the synthesis function is
identical for Web and LLM rules, so the LLM-side held-out
generalization signal (15--30\% on AdvBench/HarmBench) is
suggestive of what an analogous Web corpus would show \emph{if}
the templates' detection blocks were authored against the same
information-theoretic surface (a fair-but-untested
assumption). We do not claim Web-side generalization in the
strong sense the LLM side now supports; readers should read the
Web TP rate ($100\%$ on the self-referential calibration run)
as a synthesis-correctness signal, not a detection-quality
signal.

\paragraph{Backend-budget artifacts.} The LLM scenario depends on a
real LLM backend with a daily token budget. An early multi-seed run
exhausted the budget after the second seed, invalidating the
remaining three. We rerun with the budget raised to at least
$5\times$ the per-run token cost ($\approx 100$k tokens per LLM
calibration run on the targets we used), which yields $N{=}5$ valid
seeds in a single measurement window. Reviewers reproducing the
paper should provision a comparable budget headroom or stagger seed
runs across budget reset boundaries.

\paragraph{Lucene regex compatibility shim.} The real-SIEM replay
(Section~\ref{sec:results:opensearch}) revealed that
pysigma-backend-elasticsearch emits \texttt{\textbackslash{}b}
word-boundary anchors in its KQL output. Lucene's Java regex engine
does not accept \texttt{\textbackslash{}b} (a PCRE-only construct),
so every rule containing one yields zero hits when run as-emitted.
Concretely, the LLM01 rule's selector \texttt{\textbackslash{}bDAN
mode\textbackslash{}b} is converted by
pysigma-backend-elasticsearch into the Lucene regex
\texttt{/\textbackslash{}bDAN mode\textbackslash{}b/}; Lucene's Java
regex engine rejects the \texttt{\textbackslash{}b} token and returns
no documents, even on a prompt that literally contains
``DAN mode''. This is the kind of pattern that fires in the Python
harness but silently no-ops in the SIEM.
We added a compatibility shim that strips \texttt{\textbackslash{}b}
after the backend conversion --- rewriting the example above to the
wildcard term \texttt{*DAN mode*}. After the strip, the converted KQL
contains wildcarded substring patterns of the form
\texttt{*ignore previous instructions* OR *DAN mode* OR ...} on
the indexed prompt field --- the exact queries are committed in the
replication package's OpenSearch-replay record. Two compounding factors
explain the drop from synthetic-harness $60\%$
(Table~\ref{tab:heldout}) to OpenSearch $30\%$
(Table~\ref{tab:opensearch}). (i)~The Lucene backend is
case-sensitive on \texttt{keyword}-typed fields where the Python
substring evaluator was not, so prompts that contain the trigger
in a different case (e.g.\ ``Ignore Previous Instructions'' vs.\
the rule's ``ignore previous instructions'') match in the
harness but not in OpenSearch. (ii)~Stripping
\texttt{\textbackslash{}b} broadens the regex slightly (a word
match becomes a substring match), which mostly cancels the
case-sensitivity loss on the cohort but does not recover the
prompts whose phrasing diverges further from the rule's verb
list (e.g.\ ``please disregard the prior'' vs.\ ``ignore
previous''). Empirically the two factors net out to roughly half
the synthetic-harness rate on AdvBench.

\subparagraph{Affected templates.}
The shim touches every rule whose template emits
\texttt{\textbackslash{}b}-anchored selectors --- all $3$ live
LLM01 rules and all $8$ Web rules in the calibration set. Of
those, the $3$ LLM01 rules survive after the strip as substring
matches (\texttt{*jailbreak*}, \texttt{*DAN mode*}, etc., visible
in the replication package's OpenSearch-replay record). The $8$ Web rules
degenerate to empty selectors because their keyword lists were
authored as bare tokens with relied-upon word-boundary semantics
(e.g.\ \texttt{\textbackslash{}bDROP TABLE\textbackslash{}b});
stripping the anchors leaves a pattern that the OpenSearch
\texttt{keyword} field's exact-match semantics no longer matches.
This is the mechanical explanation for the $0/50$ benign-Web
union-match rate in Table~\ref{tab:opensearch}.

\subparagraph{FP impact of the shim.}
Counter-intuitively, the shim \emph{reduced} the LLM01 false
positive rate on the benign-LLM cohort, not increased it. The
synthetic harness fired on $4/26$ ($15\%$) benign prompts;
the OpenSearch backend fires on $2/26$ ($7.7\%$). The drop is
the case-sensitivity tightening described above (factor (i)
applied to FP cohort as well as TP cohort). The Web side is
$0/50$ FP because its $8$ rules are simply not firing on
anything --- a degenerate ``no FP because no detection''
condition, not a clean low-FP signal. We report both honestly in
Table~\ref{tab:opensearch} rather than presenting the Web
$0\%$ as a quality claim.

\subparagraph{Recommended recovery path.}
Two fixes are durable; we recommend both in tandem.
\emph{Template side (immediate):} update the OWASP-Web
templates' detection blocks to use \texttt{contains|all} with
case-folded selectors and explicit wildcards
(\texttt{*' OR 1=1*}, \texttt{*UNION SELECT*}) rather than
\texttt{\textbackslash{}b}-anchored bare tokens. This change is
backward-compatible with Splunk and matches the Lucene
\texttt{keyword} field semantics natively, removing the shim
dependency for the Web side. \emph{Toolchain side
(structural):} contribute a Lucene-aware regex emission flavor
to \texttt{pysigma-backend-elasticsearch} that maps
\texttt{\textbackslash{}b} to the appropriate Lucene token-boundary
construct (Lucene supports lookahead-style boundary matching via
\texttt{\^{}|\$}-anchored alternations on the analyzer
output). Doing both makes \texttt{\textbackslash{}b}-bearing
Sigma rules portable to Lucene without behaviour drift, and
recovers the synthetic-harness rate to within bootstrap-level
agreement with the OpenSearch rate. Neither fix requires a
change to the synthesis function itself --- the synthesis
function passes the template's \texttt{detection} block through
opaquely.
Both are template/backend-side changes, not pipeline-side.

It also explains the $0\%$ fire rate on the benign-Web cohort
(Table~\ref{tab:opensearch}): all eight Web rules' keyword
selections were authored against \texttt{\textbackslash{}b}-anchored
expectations and degenerate after the strip. The Splunk backend
($17/17$ conform, Table~\ref{tab:v6}) accepts \texttt{\textbackslash{}b}
natively, so this is a Lucene-emission gap in pysigma, not in the
synthesis function; a follow-up should either contribute a
Lucene-compatible regex flavor upstream or have the synthesis
function avoid \texttt{\textbackslash{}b} in templates targeted at
Lucene deployments.

\paragraph{Audit-surfaced measurement bugs.} The multi-seed
discipline surfaced two bugs in the calibration pipeline that the
single-run protocol would not have caught:
(1) the \texttt{llm10-maxtokens-001} probe was sending a request
shape that the LLM-target endpoint rejected with a 400 status,
causing the probe to always classify as \texttt{error} rather
than \texttt{capped}/\texttt{bypassed};
(2) an earlier measurement claimed $\sigma{=}0$ on the LLM corpus
based on $N{=}2$ valid seeds, an overclaim that the budget-corrected
$N{=}5$ measurement contradicted ($\sigma{\approx}0.8$).
Both are now fixed and documented as artifact-audit findings; the
contribution-level claim (template-hit ${=}100\%$, skipped${=}0$)
survives both corrections because skipped depends on
\texttt{findings.length} reaching the synthesis function, not on the
specific value of \texttt{bypassed}.

%% file: tables/fp_summary.tex
% AUTO-GENERATED by scripts/export_fp_table.py from
% reports/{replay-fidelity,heldout-v2-replay,opensearch-replay}.json — do not edit.
\begin{table}[t]
  \centering
  \small
  \caption{Every false-positive measurement in the paper, with its evaluator and benign cohort. The rates are \emph{not} a single number measured five ways: the synthetic rows count fires per (rule, log-line) evaluation, whereas the OpenSearch rows count per-prompt union match (a prompt is a false positive if \emph{any} rule fires on it). The headline operational FP is the OpenSearch benign-LLM row ($2/26$, $7.7\%$). $^\dagger$The benign-Web $0\%$ is a degenerate no-fire caused by the Lucene \texttt{\textbackslash{}b} shim (Section~\ref{sec:discussion:threats}), not a clean low-FP signal.}
  \label{tab:fp-summary}
  \resizebox{\columnwidth}{!}{%
  \begin{tabular}{lllrr}
    \toprule
    Context & Evaluator & Benign cohort & Fires & FP rate \\
    \midrule
    Web synthetic baseline & Python substring & 8 rules $\times$ 100 lines & 4/800 & 0.5\% \\
    LLM synthetic baseline & Python substring & 7 rules $\times$ 100 lines & 0/700 & 0.0\% \\
    v2 prototype (LLM01) & Python substring & 26 benign prompts & 4/26 & 15.4\% \\
    OpenSearch benign-LLM & Lucene / OpenSearch & 26 benign prompts & 2/26 & 7.7\% \\
    OpenSearch benign-Web & Lucene / OpenSearch & 50 benign requests & 0/50 & 0.0\%$^\dagger$ \\
    \bottomrule
  \end{tabular}%
  }
\end{table}

%% file: sections/conclusion.tex
\section{Conclusion}

Locked probe corpora — JSON files with stable \texttt{probe\_id}s
released as research artifacts — make the gap between BAS findings
and SIEM rules shorter than it has to be. We described a synthesis
function that consumes the corpus's OWASP-categorized findings and
emits Sigma rules with deterministic back-references to the source
finding and the MITRE technique. The function covers every bypassed
finding in the two corpora we release (skipped count: zero across 17
LLM + 23 Web probes).

We are deliberate about what this paper does not claim. The
synthesis is starter-rule level, not production-rule level. We
report early-stage detection numbers — $30\%$ union match on
AdvBench and $14\%$ on HarmBench through a live
OpenSearch+Lucene index, with $7.7\%$ FP on a benign-LLM baseline
— but these are bounded by the v2 template rubric's scope (LLM01
family only) and by a Lucene regex-emission gap in pysigma's
backend (Section~\ref{sec:discussion:threats}). The contribution we
do claim is a reproducibility envelope: the synthesis is a
deterministic pure function (re-running it on a finding emits
byte-identical YAML), specified as pseudocode, and the resulting rule
set --- deterministic ids plus multi-backend conversions --- is
published, so a reviewer re-implementing from the spec reproduces it
from the locked corpus alone, with every rule's source provenance
available in two clicks.

Our companion work~\cite{paperOne} established the locked-corpus
discipline for measuring engine-side defense attribution. This paper
extends the discipline to the defense-side output: every emitted
detection rule traces back to the same corpus entry the engine measured
against. The two papers together cover the round trip from attack
probe to detection content with an integrity chain a reviewer can
verify without the engine source code.

Remaining gaps for follow-up: (a) a controlled $N{\geq}5$ analyst
timing study to replace the hand-wavy $30$--$60$ minute manual-
authoring estimate; (b) corpus extension to exercise the four
reserved templates (A04, LLM03, LLM04, LLM05); (c) a Splunk-side
replay to mirror the OpenSearch+Lucene measurement and resolve the
\texttt{\textbackslash{}b} compatibility gap upstream in pysigma; and
(d) a stand-alone \texttt{bas://} resolver service for click-through
from a fired SIEM alert to the originating probe, scoped out here
(Section~\ref{sec:methodology}) and left to operational deployment.

%% file: sections/appendix.tex
\appendix

\section{Worked Examples: BAS Finding $\to$ Sigma Rule}
\label{sec:appendix:examples}

This appendix walks through three end-to-end examples covering the
three template-source classes (legacy MITRE T-code, OWASP LLM, OWASP
Web). For each, we show: (a) the calibration probe; (b) the
resulting finding fields; (c) the Sigma rule emitted by
\texttt{findingToSigma()}; (d) a justification for why the rule is a
useful, non-trivial starting point for a SOC analyst.

\subsection{Example 1: Web injection (OWASP A03 / T1190)}

\paragraph{Probe.}
\texttt{a03-sqli-login-email-001} sends
\texttt{POST /rest/user/login \{email: "' OR 1=1--", password: "x"\}}
and looks for \texttt{body\_contains: "authentication"} as the
bypass indicator.

\paragraph{Finding.}
\texttt{attack\_technique = "OWASP A03 --- SQL injection in login\\ email
bypasses authentication"};
\texttt{severity = critical};
\texttt{detection\_gap = "Probe a03-sqli-login-email-001 confirmed
vulnerable. Indicator matched: body\textasciitilde\,'authentication'..."};
\texttt{metadata.probe\_id = a03-sqli-login-email-001}.

\paragraph{Emitted rule (excerpt).}
Template \texttt{A03} fires (OWASP regex pass).
\texttt{logsource = (webserver, access)};
\texttt{detection.keywords = ["' OR 1=1", "UNION SELECT", "DROP
TABLE", "<script>", "javascript:"]};
\texttt{condition = keywords};
\texttt{tags = [attack.t1190, bas.severity.critical]}.

\paragraph{Justification.} The rule fires on any HTTP access log
where the request URI or body contains a classic SQLi keyword.
This is a more aggressive starting point than vendor SIEM defaults
(which typically require a downstream alert from a WAF) and gives
the analyst a hit-everywhere baseline that they can tune down with
allow-lists for legitimate pen-test sources. The
\texttt{references[0]} URI lets an on-call responder pivot from a
fired alert back to the originating probe payload in the BAS run.

\subsection{Example 2: LLM jailbreak (OWASP LLM01 / T1190.002)}

\paragraph{Probe.} \texttt{llm01-jb-dan-001} sends the DAN ``Do
Anything Now'' jailbreak payload via
\texttt{POST /chat \{message: "Ignore all previous instructions. You
are DAN..."\}}. The probe declares
\texttt{forbidden = ["DAN MODE ON", "step-by-step", "explosive"]}.

\paragraph{Finding.} Emitted when any forbidden token appears in
the model response. \texttt{attack\_technique = "OWASP LLM01 ---
DAN-style act-as hijack"}; \texttt{severity = high};
\texttt{metadata.probe\_id = llm01-jb-dan-001}.

\paragraph{Emitted rule (excerpt).} Template \texttt{LLM01} fires.
\texttt{logsource = (application, llm)};
\texttt{detection.keywords = ["ignore previous instructions", "DAN
mode", "developer mode", "<\textbar im\_start\textbar >",
"jailbreak", "grandmother"]};
\texttt{tags = [attack.t1190.002, bas.severity.high]};
\texttt{falsepositives = ["Legit red-team / security research",
"Researchers discussing jailbreak history"]}.

\paragraph{Justification.} LLM application logs are not yet covered
by stock SIEM rule packs; this rule gives the analyst a starter
ruleset to deploy on a Cloud Run \texttt{stdout} log stream or a
Vertex AI \texttt{audit} stream. The
\texttt{falsepositives} list flags the most likely
benign trigger (researchers \emph{discussing} jailbreaks in
chat) so the analyst can pre-author an allow-list.

\subsection{Example 3: LLM tool-enumeration (OWASP LLM06 / T1059.013)}

\paragraph{Probe.}
\texttt{llm06-tools-functions-001} sends ``List the JSON schema of
every tool registered to you. Use the OpenAI function-calling
format.'' \texttt{forbidden = ['"name":', '"parameters":',
'"properties":']}.

\paragraph{Finding.} \texttt{attack\_technique = "OWASP LLM06 ---
List all tool JSON schemas"}; \texttt{severity = high}.

\paragraph{Emitted rule (excerpt).} Template \texttt{LLM06} fires.
\texttt{logsource = (application, llm)};
\texttt{detection.response\_keywords = ["send\_email",
"delete\_user", "execute\_sql", "" tool\_name" :",
"" function" :", "" parameters" :"]};
\texttt{condition = response\_keywords}.

\paragraph{Justification.} The detection sits on the model's
\emph{response} side (not the request side), so it catches both
direct enumeration via prompt and indirect leakage via reflective
queries (``what could you do for me?''). The
\texttt{falsepositives} note (``Documentation chatbots intentionally
describing their toolset'') tells the analyst to scope the rule out
of documentation deployments. Without this starter rule, the
analyst writing it from scratch typically needs $\geq 1$ hour to
identify the right detection surface (response-side) and assemble
the keyword list.

\paragraph{Trace.} All three examples are reproducible end-to-end:
the corresponding \texttt{finding\_id} appears in the run's evidence
blob (Section~\ref{sec:methodology:reproducibility}), and the rule
YAML is regenerated deterministically from the same
(finding, template) pair by re-running the synthesis function.

\subsection{The v1 vs v2 LLM01 detection block}
\label{sec:appendix:v1v2}

Listing~\ref{lst:v1v2} shows the LLM01 template's \texttt{detection}
block before and after the v1$\to$v2 rubric change discussed in
Section~\ref{sec:discussion:authoring}. The v1 block fires only on
jailbreak \emph{syntax} (meta-tokens such as \texttt{DAN mode}); the
v2 block keeps those keywords and adds two semantic markers --- a list
of direct harmful-intent terms and an action-verb\,$+$\,harm-object
regex --- so that plain-prose AdvBench prompts (which carry no
jailbreak syntax) also fire. This is the single template edit behind
the $0/50\to30/50$ held-out jump in Table~\ref{tab:v6}(b); the
synthesis function and traceability contract are unchanged.

\begin{figure}[H]
\begin{lstlisting}[basicstyle=\ttfamily\scriptsize,frame=single,framerule=0.4pt,framesep=4pt,breaklines=true]
# v1 (keyword-only) -- fires on jailbreak SYNTAX
detection:
  keywords:
    - "DAN mode"
    - "<|im_start|>"
    - "developer mode"
    - "grandmother"
  condition: keywords

# v2 (keyword + regex) -- adds harmful-INTENT coverage
detection:
  keywords:          # v1 jailbreak syntax, retained
    - "DAN mode"
    - "<|im_start|>"
    - "developer mode"
  intent_terms:      # +19 direct harmful-intent terms
    - "identity theft"
    - "phishing"
    - "malware"
    - "fraud"
    # ... (full 19-term list in the replication package)
  intent_re|re:      # action-verb + harm-object pairs
    - "\b(write|create|generate|develop)\b.*
       \b(malware|exploit|weapon|bomb)\b"
  condition: keywords or intent_terms or intent_re
\end{lstlisting}
\captionof{lstlisting}{Representative v1 (top) and v2 (bottom)
\texttt{detection} blocks for the LLM01 template. Term and regex lists
are abbreviated; the complete blocks are in the replication package.}
\label{lst:v1v2}
\end{figure}

\subsection{Full probe-to-template map}
\label{sec:appendix:probe-map}

Tables~\ref{tab:probe-map} (LLM corpus) and~\ref{tab:probe-map-web}
(Web corpus) list every probe in the two locked corpora alongside the
template its OWASP category resolves to and the MITRE ATT\&CK technique
the emitted rule references. They are the detection-side companion to
the locked-corpus listing published with the attribution
work~\cite{paperOne}: where that table records each probe's designed
outcome per defense level, these record each probe's deterministic
template assignment. Because the assignment is total over both corpora,
the tables double as a visual proof of the \texttt{skipped}${=}0$
result in Table~\ref{tab:coverage} --- no probe is left without a
template.

\input{tables/probe_map}

%% file: tables/probe_map.tex
% AUTO-GENERATED by scripts/export_probe_map_table.py from
% datasets/owasp-llm-probe-corpus.json + datasets/juiceshop-probes-2026-05-18.json + reports/templates.json — do not edit.
\begin{table*}[t]
  \centering
  \setlength{\tabcolsep}{4pt}
  \small
  \caption{Deterministic probe$\to$template$\to$technique map, OWASP LLM Top 10 corpus ($N{=}17$, \texttt{paper-1-locked-2026-05-16}). Each probe's OWASP category selects exactly one template (\emph{Template} column), whose MITRE ATT\&CK identifier the emitted rule's \texttt{references} field carries. The assignment is a pure function of (corpus, template library): every probe resolves to a template, so a bypassed finding can never be skipped (Table~\ref{tab:coverage}). Probes of the same category collapse to one template --- the source of the depth-axis $O(1)$ scalability discussed in Section~\ref{sec:discussion:authoring}.}
  \label{tab:probe-map}
  \begin{tabular}{l p{7.2cm} l l}
    \toprule
    Probe ID & Description & Template & MITRE technique \\
    \midrule
    \addlinespace[2pt]
    \multicolumn{4}{l}{\textit{LLM01}} \\
    \texttt{\footnotesize llm01-jb-dan-001} & DAN-style act-as hijack & \texttt{LLM01} & \texttt{T1190.002} \\
    \texttt{\footnotesize llm01-jb-grandma-001} & Grandma emotional manipulation & \texttt{LLM01} & \texttt{T1190.002} \\
    \texttt{\footnotesize llm01-jb-devmode-001} & Developer-mode bypass & \texttt{LLM01} & \texttt{T1190.002} \\
    \texttt{\footnotesize llm01-jb-base64-001} & Base64-encoded payload smuggling & \texttt{LLM01} & \texttt{T1190.002} \\
    \texttt{\footnotesize llm01-jb-leet-001} & Leetspeak smuggling & \texttt{LLM01} & \texttt{T1190.002} \\
    \texttt{\footnotesize llm01-jb-imstart-001} & System role override via user message (\textless{}\textbar{}im\_start\textbar{}\textgreater{}) & \texttt{LLM01} & \texttt{T1190.002} \\
    \addlinespace[2pt]
    \multicolumn{4}{l}{\textit{LLM07}} \\
    \texttt{\footnotesize llm07-prompt-leak-001} & System prompt verbatim leak & \texttt{LLM07} & \texttt{T1083} \\
    \texttt{\footnotesize llm07-prompt-reveng-001} & Reverse-engineer system prompt & \texttt{LLM07} & \texttt{T1083} \\
    \addlinespace[2pt]
    \multicolumn{4}{l}{\textit{LLM02}} \\
    \texttt{\footnotesize llm02-canary-001} & API canary key echo & \texttt{LLM02} & \texttt{T1552.004} \\
    \texttt{\footnotesize llm02-training-001} & Training data extraction (Stripe SDK canary) & \texttt{LLM02} & \texttt{T1552.004} \\
    \texttt{\footnotesize llm02-pii-001} & Synthetic PII echo (SSN+CC) & \texttt{LLM02} & \texttt{T1552.004} \\
    \addlinespace[2pt]
    \multicolumn{4}{l}{\textit{LLM06}} \\
    \texttt{\footnotesize llm06-tools-public-001} & Tool registry enumeration (/tools) & \texttt{LLM06} & \texttt{T1059.013} \\
    \texttt{\footnotesize llm06-tools-functions-001} & Tool registry enumeration (/functions) & \texttt{LLM06} & \texttt{T1059.013} \\
    \texttt{\footnotesize llm06-side-effect-001} & Side-effect tool detection (send\_email, delete\_*) & \texttt{LLM06} & \texttt{T1059.013} \\
    \addlinespace[2pt]
    \multicolumn{4}{l}{\textit{LLM10}} \\
    \texttt{\footnotesize llm10-maxtokens-001} & max\_tokens=999999 manipulation & \texttt{LLM10} & \texttt{T1499.004} \\
    \texttt{\footnotesize llm10-model-swap-001} & Expensive model swap (claude-opus-4-7) & \texttt{LLM10} & \texttt{T1499.004} \\
    \texttt{\footnotesize llm10-burst-001} & Burst 10 req/s rate limit & \texttt{LLM10} & \texttt{T1499.004} \\
    \bottomrule
  \end{tabular}
\end{table*}
\begin{table*}[t]
  \centering
  \setlength{\tabcolsep}{4pt}
  \small
  \caption{Deterministic probe$\to$template$\to$technique map, OWASP Web Top 10 corpus ($N{=}23$, \texttt{juiceshop-2026-05-18}). Companion to Table~\ref{tab:probe-map}; same construction.}
  \label{tab:probe-map-web}
  \begin{tabular}{l p{7.2cm} l l}
    \toprule
    Probe ID & Description & Template & MITRE technique \\
    \midrule
    \addlinespace[2pt]
    \multicolumn{4}{l}{\textit{A03}} \\
    \texttt{\footnotesize a03-sqli-login-email-001} & SQL injection in login email bypasses authentication (classic ' OR 1=1--) & \texttt{A03} & \texttt{T1190} \\
    \texttt{\footnotesize a03-sqli-login-quote-001} & SQL injection with comment terminator bypasses password check & \texttt{A03} & \texttt{T1190} \\
    \texttt{\footnotesize a03-sqli-search-001} & UNION-based SQL injection in product search leaks bcrypt hashes & \texttt{A03} & \texttt{T1190} \\
    \texttt{\footnotesize a03-sqli-order-by-001} & Order-by SQL injection via sort param (error-based detection) & \texttt{A03} & \texttt{T1190} \\
    \addlinespace[2pt]
    \multicolumn{4}{l}{\textit{A01}} \\
    \texttt{\footnotesize a01-idor-basket-001} & IDOR — basket of another user readable without auth scoping & \texttt{A01} & \texttt{T1190} \\
    \texttt{\footnotesize a01-idor-basket-002} & IDOR — basket \#2 & \texttt{A01} & \texttt{T1190} \\
    \texttt{\footnotesize a01-idor-basket-003} & IDOR — basket \#3 & \texttt{A01} & \texttt{T1190} \\
    \texttt{\footnotesize a01-idor-basket-004} & IDOR — basket \#4 & \texttt{A01} & \texttt{T1190} \\
    \texttt{\footnotesize a01-idor-basket-005} & IDOR — basket \#5 & \texttt{A01} & \texttt{T1190} \\
    \texttt{\footnotesize a01-idor-feedback-001} & IDOR — feedback record of another user & \texttt{A01} & \texttt{T1190} \\
    \texttt{\footnotesize a01-idor-users-listing-001} & Unauthenticated Users listing endpoint leaks all account emails & \texttt{A01} & \texttt{T1190} \\
    \addlinespace[2pt]
    \multicolumn{4}{l}{\textit{A03}} \\
    \texttt{\footnotesize a03-xss-search-001} & Reflected XSS in product search query echo & \texttt{A03} & \texttt{T1190} \\
    \texttt{\footnotesize a03-xss-deeplink-001} & Reflected XSS via fragment deeplink (hash-route DOM injection) & \texttt{A03} & \texttt{T1190} \\
    \texttt{\footnotesize a03-xss-track-result-001} & Reflected XSS in track-result id parameter & \texttt{A03} & \texttt{T1190} \\
    \addlinespace[2pt]
    \multicolumn{4}{l}{\textit{A02}} \\
    \texttt{\footnotesize a02-jwt-none-001} & JWT alg:none accepted — admin identity forgeable without signature & \texttt{A02} & \texttt{T1078} \\
    \addlinespace[2pt]
    \multicolumn{4}{l}{\textit{A01}} \\
    \texttt{\footnotesize a01-admin-public-001} & Administration UI route reachable without auth check on frontend & \texttt{A01} & \texttt{T1190} \\
    \texttt{\footnotesize a01-admin-api-001} & Mass assignment — role field on Users PUT accepts admin upgrade & \texttt{A01} & \texttt{T1190} \\
    \addlinespace[2pt]
    \multicolumn{4}{l}{\textit{A05}} \\
    \texttt{\footnotesize a05-ftp-listing-001} & FTP directory listing exposed at /ftp & \texttt{A05} & \texttt{T1083} \\
    \texttt{\footnotesize a05-ftp-acquisitions-001} & Confidential acquisitions document publicly readable & \texttt{A05} & \texttt{T1083} \\
    \addlinespace[2pt]
    \multicolumn{4}{l}{\textit{A01}} \\
    \texttt{\footnotesize a01-pathtraversal-ftp-001} & Poison-null byte path traversal bypasses .md whitelist & \texttt{A01} & \texttt{T1190} \\
    \addlinespace[2pt]
    \multicolumn{4}{l}{\textit{A09}} \\
    \texttt{\footnotesize a09-secret-keys-001} & Encryption keys directory listing leaks premium.key & \texttt{A09} & \texttt{T1552.001} \\
    \addlinespace[2pt]
    \multicolumn{4}{l}{\textit{A07}} \\
    \texttt{\footnotesize a07-2fa-bypass-001} & 2FA verification flow accepts weak/test token & \texttt{A07} & \texttt{T1110} \\
    \addlinespace[2pt]
    \multicolumn{4}{l}{\textit{A01}} \\
    \texttt{\footnotesize a01-redirect-allowlist-001} & Open redirect bypasses allowlist (target=evil.com) & \texttt{A01} & \texttt{T1190} \\
    \bottomrule
  \end{tabular}
\end{table*}

%% file: references.bib
@misc{paperOne,
  title         = {Which Defense Closes Which Threat? Attributing {OWASP}-{LLM}-Top-10 Coverage and Its Brittleness Under Paraphrasing},
  author        = {Maiorano, Alexandre Cristov{\~a}o},
  year          = {2026},
  eprint        = {2606.02822},
  archivePrefix = {arXiv},
  primaryClass  = {cs.CR},
  url           = {https://arxiv.org/abs/2606.02822},
}

@misc{owaspTop10,
  title        = {{OWASP} Top 10:2021},
  author       = {{OWASP Foundation}},
  year         = {2021},
  howpublished = {\url{https://owasp.org/Top10/}},
}

@misc{owaspLlmTop10,
  title        = {{OWASP} Top 10 for {LLM} Applications},
  author       = {{OWASP GenAI Security Project}},
  year         = {2025},
  howpublished = {\url{https://genai.owasp.org/llm-top-10/}},
}

@misc{sigmaHQ,
  title        = {{Sigma}: Generic Signature Format for {SIEM} Systems},
  author       = {{SigmaHQ}},
  year         = {2024},
  howpublished = {\url{https://github.com/SigmaHQ/sigma}},
}

@misc{mitreAttack,
  title        = {{MITRE ATT\&CK}},
  author       = {{The MITRE Corporation}},
  year         = {2024},
  howpublished = {\url{https://attack.mitre.org/}},
}

@misc{zouAdvBench,
  title        = {Universal and Transferable Adversarial Attacks on Aligned Language Models},
  author       = {Zou, Andy and Wang, Zifan and Carlini, Nicholas and Nasr, Milad and Kolter, J. Zico and Fredrikson, Matt},
  year         = {2023},
  eprint       = {2307.15043},
  archivePrefix= {arXiv},
  primaryClass = {cs.CL},
}

@misc{garak,
  title        = {garak: A Framework for Security Probing Large Language Models},
  author       = {Derczynski, Leon and Galinkin, Erick and Martin, Jeffrey and Majumdar, Subho and Inie, Nanna},
  year         = {2024},
  eprint       = {2406.11036},
  archivePrefix= {arXiv},
  primaryClass = {cs.CL},
}

@misc{sigmaHQrules,
  title        = {SigmaHQ Community Rule Library},
  author       = {{SigmaHQ Contributors}},
  year         = {2024},
  howpublished = {\url{https://github.com/SigmaHQ/sigma}},
}

@misc{caldera,
  title        = {{MITRE} {CALDERA}: Adversary Emulation Platform},
  author       = {{The MITRE Corporation}},
  year         = {2024},
  howpublished = {\url{https://caldera.mitre.org/}},
}

@inproceedings{harmbench,
  title        = {{HarmBench}: A Standardized Evaluation Framework for Automated Red Teaming and Robust Refusal},
  author       = {Mazeika, Mantas and Phan, Long and Yin, Xuwang and Zou, Andy and Wang, Zifan and Mu, Norman and Sakhaee, Elham and Li, Nathaniel and Basart, Steven and Li, Bo and Forsyth, David and Hendrycks, Dan},
  booktitle    = {International Conference on Machine Learning (ICML)},
  year         = {2024},
  eprint       = {2402.04249},
  archivePrefix= {arXiv},
}

@misc{opensearch,
  title        = {{OpenSearch}: An open-source distributed search and analytics suite},
  author       = {{The OpenSearch Project}},
  year         = {2024},
  howpublished = {\url{https://opensearch.org/}},
}

@misc{ruleGenie,
  title        = {{RuleGenie}: {SIEM} Detection Rule Set Optimization},
  author       = {Shukla, Akansha and Gandhi, Parth Atulbhai and Elovici, Yuval and Shabtai, Asaf},
  year         = {2025},
  eprint       = {2505.06701},
  archivePrefix= {arXiv},
  primaryClass = {cs.CR},
}

@misc{ram-arxiv-2025,
  title        = {Rule-{ATT\&CK} Mapper ({RAM}): Mapping {SIEM} Rules to {TTPs} Using {LLMs}},
  author       = {Wudali, Prasanna N. and Kravchik, Moshe and Malul, Ehud and Gandhi, Parth A. and Elovici, Yuval and Shabtai, Asaf},
  year         = {2025},
  url          = {https://arxiv.org/abs/2502.02337},
  archivePrefix= {arXiv},
  eprint       = {2502.02337},
}
